\documentclass[acm,sigconf]{acmart}

\copyrightyear{2026}
\acmYear{2026}
\setcopyright{cc}
\setcctype{by}
\acmConference[ICSE '26]{2026 IEEE/ACM 48th International Conference on Software Engineering}{April 12--18, 2026}{Rio de Janeiro, Brazil}
\acmBooktitle{2026 IEEE/ACM 48th International Conference on Software Engineering (ICSE '26), April 12--18, 2026, Rio de Janeiro, Brazil}
\acmPrice{}
\acmDOI{10.1145/3744916.3773272}
\acmISBN{979-8-4007-2025-3/2026/04}

\AtBeginDocument{%
  }

\usepackage{amsmath,amsfonts}
\usepackage{pifont}
\usepackage{hyperref}
\usepackage{textcomp}
\usepackage{fancyhdr}
\usepackage{url}
\usepackage{cleveref}
\usepackage[dvipsnames]{xcolor}
\usepackage[table]{xcolor}
\usepackage[noend]{algpseudocode}
\usepackage{graphicx}
\usepackage{textcomp}
\usepackage{xcolor}
\usepackage{enumitem}
\usepackage{soul}
\usepackage[belowskip=0pt,aboveskip=5pt]{subcaption}
\usepackage{listings}
\usepackage{minted}
\usepackage{xspace}
\usepackage[most]{tcolorbox}
\usepackage{multirow}
\usepackage{color}
\usepackage{url}
\usepackage{cleveref}
\usepackage[export]{adjustbox}
\usepackage[linesnumbered,ruled,vlined]{algorithm2e}
\usepackage{wrapfig}
\usepackage{multirow}
\usepackage{makecell}
\crefformat{section}{\S#2#1#3} 
\crefformat{subsection}{\S#2#1#3}
\crefformat{subsubsection}{\S#2#1#3}
\usepackage{stmaryrd}
\usepackage{bm}
\usepackage{diagbox}
\definecolor{codegreen}{rgb}{0,0.6,0}
\definecolor{codegray}{rgb}{0.5,0.5,0.5}
\definecolor{codepurple}{rgb}{0.58,0,0.82}
\definecolor{backcolour}{rgb}{0.95,0.95,0.92}

\let\cite\citep
\lstdefinestyle{stylewithcommentPy}{
language=Python,
  basicstyle=\linespread{0.9}\footnotesize,
  backgroundcolor=\color{backcolour},
  commentstyle=\color{codegreen},
  keywordstyle=\color{magenta},
  numberstyle=\tiny\color{codegray},
  basicstyle=\scriptsize\ttfamily,
  escapechar=|,
  stringstyle=\color{codepurple}, xleftmargin=0cm,
  frame=tlbr, framesep=0.1cm, framerule=0pt, numbers=none,
  breaklines=true,
    moredelim=**[is][\color{red}]{@}{@},
    moredelim=**[is][\color{black}]{`}{`},
    moredelim=[is][\color{red}\bfseries]{~}{~},
    moredelim={[is][\color{black}\bfseries]{@@}{@@}}
}

\newcommand{\approach}{\textsc{CES}\xspace}
\newcommand{\codemind}{\textsc{CodeMind}\xspace}
\newcommand{\crux}{\textsc{CruxEval}\xspace}
\newcommand{\reval}{\textsc{REval}\xspace}
\newcommand{\heval}{\textsc{HumanEval}\xspace}
\newcommand{\coco}{\textsc{CocoNut}\xspace}

\newcommand{\codellama}{CodeLlama\xspace}
\newcommand{\deepseek}{DeepSeekCoder\xspace}
\newcommand{\starT}{StarCoder2\xspace}
\newcommand{\gemini}{Gemini-1.5-Pro\xspace}
\newcommand{\gptf}{GPT-4\xspace}
\newcommand{\magic}{MagiCoder\xspace}
\newcommand{\sem}{SemCoder\xspace}
\newcommand{\omini}{o4-mini\xspace}
\newcommand{\dsr}{DeepSeek-R1\xspace}

\newcommand{\correct}{confident success\xspace}
\newcommand{\sus}{suspicious success\xspace}
\newcommand{\lesssus}{likely success\xspace}

\definecolor{correctout}{RGB}{217, 234, 211}
\definecolor{incorrectout}{RGB}{244, 204, 204}
\definecolor{correctres}{RGB}{254, 248, 230}
\definecolor{incorrectres}{RGB}{252, 221, 188}

\newcommand{\rebuttalColor}{black}
\newcommand{\MRColor}{black}

\newcommand{\CRColor}{black}

\NewDocumentCommand{\reyhan}{mO{}}{\textcolor{magenta}{\textsuperscript{\textit{Reyhan}}\textsf{\textbf{\small[#1]}}}}

\AtBeginDocument{%
  }

\begin{document}

\title{Assessing Coherency and Consistency of Code Execution Reasoning by Large Language Models}


\author{Changshu Liu, Yang Chen, Reyhaneh Jabbarvand}
\email{{cl144,yangc9,reyhaneh}@illinois.edu}
\affiliation{%
  \institution{University of Illinois Urbana-Champaign}
  \city{Urbana}
  \state{Illinois}
  \country{USA}
}

\begin{abstract}
\textcolor{\rebuttalColor}{This paper proposes \approach (\textbf{C}ode \textbf{E}xecution \textbf{S}imulation), a task to evaluate the abilities of LLMs in simulating program execution}. Besides measuring the correctness of variable predictions during execution simulation, \approach introduces the notion of \textbf{coherence} to determine whether the simulation complies with commonsense execution logic, even if the predicted values along the simulations are incorrect. This enables \approach to rule out suspiciously correct output predictions due to reasoning shortcuts, hallucinations, or potential data leakage. \approach also introduces a novel metric to measure reasoning \textbf{consistency} across tests with the same or different prime path coverage in a spectrum: \emph{strong}, \emph{weak}, and \emph{random}. 

Evaluating \textcolor{\rebuttalColor}{$16$} LLMs \textcolor{\rebuttalColor}{(including three reasoning LLMs)} using \approach indicates \textcolor{\rebuttalColor}{$81.42\%$} coherent execution simulation on \heval, \textcolor{\rebuttalColor}{$46.92\%$} and \textcolor{\rebuttalColor}{$53.08\%$} of which result in correct and incorrect output predictions. Frontier LLMs such as \gptf and DeepSeek-R1 have the most incoherent execution reasoning, mostly due to natural-language shortcuts. Despite relatively coherent execution simulation, LLMs' reasoning performance across different tests is inconsistent, mostly random (\textcolor{\rebuttalColor}{$48.87\%$}) or weak (\textcolor{\rebuttalColor}{$45.37\%$}), potentially explaining their weakness in programming tasks that require path-sensitive program analysis to succeed. We also compare \approach with bug prediction/localization/repair, which intuitively requires control- and data-flow awareness. We observe that LLMs rarely incorporate execution reasoning into their analysis for bug-related tasks, and their success is primarily due to inherent pattern-matching capabilities, natural language shortcuts, or data leakage. Without reasoning, there is a threat to the generalizability of LLMs in dealing with unseen bugs or patterns in different contexts. \approach can be used to vet the suspicious success of LLMs in these tasks systematically. 
\end{abstract}

\ccsdesc[500]{Software and its engineering~Software notations and tools}
\ccsdesc[500]{Software and its engineering~Correctness}

\keywords{Large Language Models, Code Reasoning, Program Analysis}
\maketitle

\vspace{-8pt}
\section{Introduction}
\label{sec:introduction}
Large Language Models (LLMs) have shown emerging abilities in code completion and synthesis, bug/vulnerability detection, code translation, and program repair. The extent to which they can reason about different aspects of code remains under investigation. \crux~\cite{gu2024cruxeval} and \codemind~\cite{liu2024codemind,liu2025exerscope} proposed the input or output prediction tasks to evaluate code reasoning of LLMs. \reval~\cite{chen2024reasoning} took one step further and evaluated LLMs using four runtime prediction tasks: for given 
statement,
predict (1) if the statement is covered during execution, (2) variable values after the execution of it, (3) the next statement executed after it, and (4) final output. 
These techniques lack the following essential features: 

\vspace{3pt}
\noindent \textbf{\textcolor{\rebuttalColor}{Reasoning Coherency}.} 
\crux and \codemind only concern the prediction of input or output without evaluating the intermediate program states. \reval evaluates runtime behavior prediction of \emph{a subset of statements} in programs, not all, due to computational overhead inherent in their design. It also prompts an LLM separately per individual statement without \emph{reliable} strategies to combine all predictions for the entire program. \textcolor{\rebuttalColor}{As a result, none of these techniques can \emph{\textcolor{\CRColor}{efficiently}\footnote{efficiency refers to manageable computational cost and inference overhead}} and \emph{reliably} determine reasoning coherency: (1) \emph{where} the execution reasoning diverges from the ground truth for incorrect output prediction cases or (2) flag \emph{suspiciously correct output predictions} (correct prediction of the output despite the divergence of execution reasoning from the ground truth). The former pinpoints the root cause of limited reasoning capabilities of LLMs and provides insights for improving the next generation of Code LLMs. The latter can reveal data contamination, hallucinations, and reasoning shortcuts to enhance trustworthiness~\cite{zhang2023siren,shidetecting,ding2024break}}.

\vspace{3pt}
\noindent \textbf{Reasoning Consistency.} Prior techniques evaluate the reasoning of LLMs per single test, failing to study the consistency of reasoning across multiple tests with different or same coverage. Investigating such consistency can reveal the strength of inductive reasoning in LLMs: a model that correctly reasons about execution across tests with different coverage is unlikely to succeed by chance (\emph{strong} consistency), and a model that correctly reasons about the execution across tests with the same coverage, but not those with difference coverage, has a \emph{weak} consistency. Otherwise, the execution reasoning of the model can be considered random.  

This paper proposes \approach, a novel task to evaluate how LLMs can simulate the execution of given inputs. \approach unifies output prediction and intermediate program state predictions into \emph{one prompt}. Asking LLM to predict all intermediate program states and output makes the task complex and can confuse it, especially since this requires reasoning across multiple statements considering execution flow~\cite{allamanisunsupervised,sabbatella2024prompt,chang2024efficient}. \approach alleviates the issue by (1) prompting the model to predict essential decision point values (loop variables, loop iterables, conditional predicates, and branches) and (2) instructing the task with adaptive in-context examples (using static analysis to adjust the in-context example based on the program). 

\approach is \emph{flow-sensitive} (evaluates the execution reasoning of a program as a whole), \emph{scalable} (prompts LLM once per program and test input), and \emph{diagnostic} \textcolor{\rebuttalColor}{(automatically determines \emph{where} the execution reasoning diverges from the ground truth for incorrect output predictions or flags \emph{suspiciously correct output predictions} despite the divergence of execution reasoning from the ground truth)}. It can also evaluate reasoning consistency across tests with the same or different coverage. These novelties enable an extensive empirical evaluation of \textcolor{\rebuttalColor}{$16$} API- and open-access LLMs \textcolor{\rebuttalColor}{(including three reasoning models)} using \approach with the following notable findings\footnote{\approach artifacts are publicly available for assessment and reproducibility purpose~\cite{website}.}:

\hspace{-6pt} $\bullet$ \hspace{1pt} LLMs, to an extent of \textcolor{\rebuttalColor}{$81.42\%$}, on average, can coherently simulate execution of \heval programs, \textcolor{\rebuttalColor}{$46.92\%$} and \textcolor{\rebuttalColor}{$53.08\%$} of which with correct and incorrect output predictions (\S \ref{subsec:rq1}). \gptf and \dsr demonstrate higher levels of incoherent reasoning due to natural language shortcuts. Despite coherent execution simulation, LLMs’ reasoning across different tests is inconsistent (\S \ref{rq:reasoning-consistency}), mostly random (\textcolor{\rebuttalColor}{$48.87\%$}), or weak (\textcolor{\rebuttalColor}{$45.37\%$}). 

\hspace{-4pt} $\bullet$ \hspace{1pt} \approach categorizes the root causes for incorrect or suspiciously correct output predictions (\S \ref{subsec:rq3}), which could be valuable to understanding the limitations of LLMs in code reasoning and designing the next generation of execution-aware Code LLMs.

\hspace{-4pt} $\bullet$ \hspace{1pt} We compare \approach with bug prediction/localization/repair tasks that succeeding in them requires control and data flow awareness in classic software analysis. We observe an overlap between the success of LLMs in those three tasks and \approach. However, \emph{even when pre-trained on execution data, LLMs barely incorporate execution reasoning into their analysis for bug-related tasks (\S \ref{subsec:rq4})}. We show that the novel design of \approach enables a \emph{systematic vetting of suspicious success in these tasks}, showing that most successes are due to lucky hallucinations or natural language shortcuts. This aligns with LLMs' observed lack of generalizability beyond benchmarks and in real-world settings. \approach is the first work that connects code reasoning with other programming tasks and offers a systematic way for reliable vetting of suspiciously correct cases.    

\vspace{-8pt}
\section{Illustrative Example}
\label{sec:example}

\begin{figure}[t]
    \centering
    \includegraphics[width=0.85\linewidth]{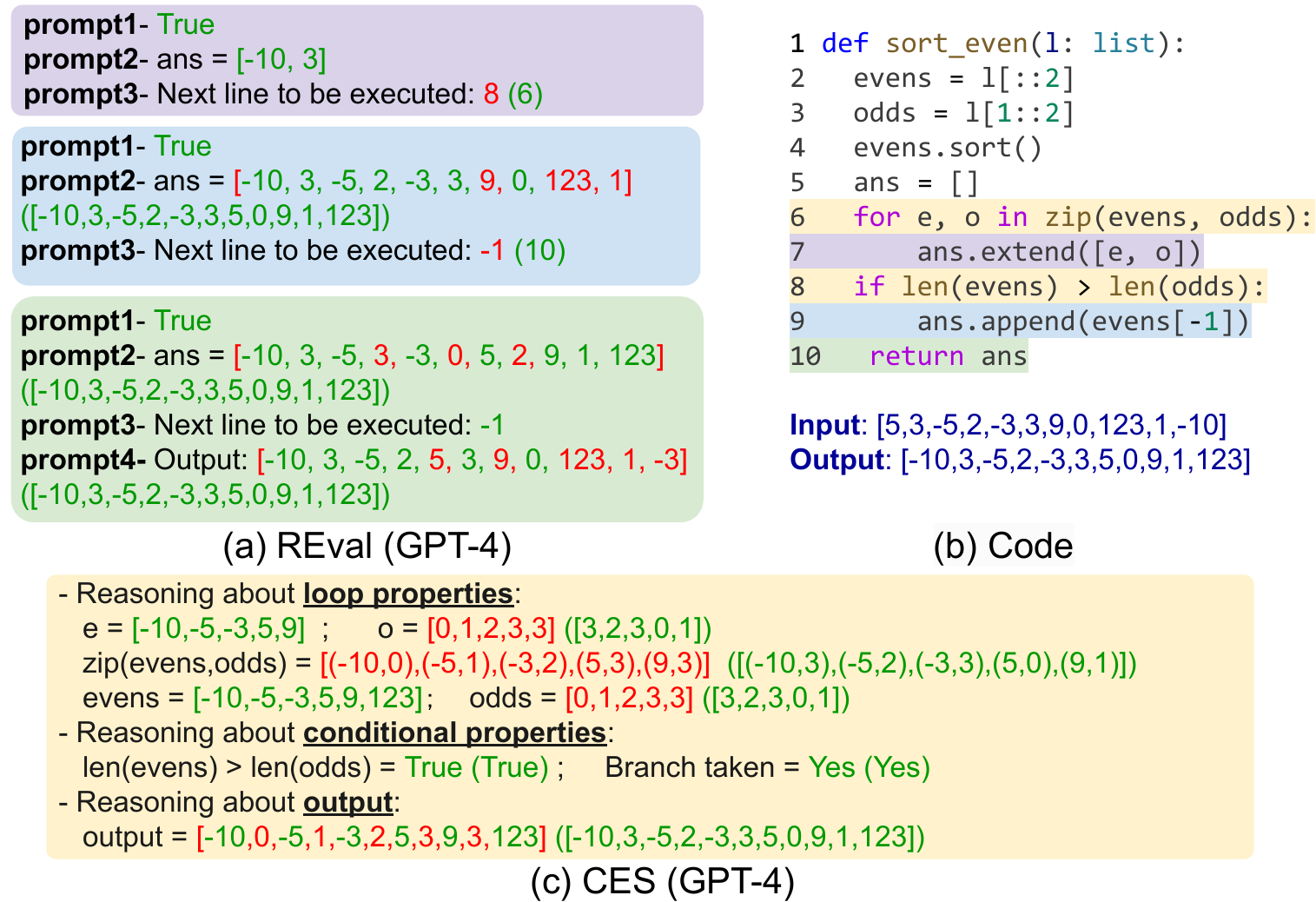}
    \vspace{-8pt}
    \caption{Output prediction of \gptf for HumanEval/37 (b); Reasoning analysis by \approach (c) and \reval (a)}
    \vspace{-22pt}
    \label{fig:illistrative-1}
\end{figure}

\textcolor{\rebuttalColor}{LLMs may not perfectly simulate the execution of the entire program, i.e., the simulation may diverge from real execution due to misprediction of some intermediate values. We refer to this phenomenon as \emph{simulation divergence}. Understanding where the model's execution simulation starts to diverge from real execution (simulation divergence point) can reveal important facts about LLMs' limitations in code reasoning (\S \ref{subsec:rq1} and \S \ref{subsec:rq3}), and whether they incorporate reasoning properly when solving other programming tasks (\S \ref{subsec:rq4}).} It is also important to \emph{rule out} suspiciously correct output predictions that can happen due to data contamination (the expected output of the program for given inputs has been seen during training), hallucinations (correct prediction based on previously incorrect ones), or shortcuts (predicting the code logic based on the function name, not understanding the code). Otherwise, the execution reasoning abilities of LLMs can be incorrectly inflated. As we will show in this paper, suspiciously correct output prediction is common, even higher for strong reasoning LLMs such as \omini and Gemini-2.5-Pro compared to other models (\cref{subsec:rq4}). \textcolor{\rebuttalColor}{This section illustrates the importance of reasoning coherence and the limitations of prior work in determining incoherent reasoning.}  
Consider Figure~\ref{fig:illistrative-1}-b, which shows the code and corresponding test for the HumanEval/37.
\textcolor{\CRColor}{For output prediction alone, \codemind, \reval, and \approach\footnote{\crux does not have evaluation results on \heval programs.} consistently evaluate GPT-4’s predictions as incorrect.}
\codemind provides no additional information on where the model lost the execution track, producing incorrect output predictions. \reval selects statements $7$, $9$, and $10$ as representative statements for code reasoning \textcolor{\rebuttalColor}{(highlighted with colors purple, blue, and green, respectively)}. For these statements, \reval individually prompts \gptf to predict if the statement is covered in the execution (prompt1), what the variable values are after the execution (prompt2), and what the next statement to be executed is (prompt3). For statement $10$, \reval asks the model to also predict the output value (prompt4). Figure~\ref{fig:illistrative-1}-a summarizes the \gptf response to $10$ \reval prompts. \textcolor{\rebuttalColor}{Prompt sets are colored to match the highlighted color of corresponding statements. Ground truth values are green, and if a model prediction is different from ground truth, the figure highlights incorrect predictions in red.} 

\textcolor{\MRColor}{Per \reval analysis, \gptf correctly responds to prompt1 (predicting coverage) for all selected statements. In response to prompt2 (predicting program state), \gptf's response is correct, predicting the values of \texttt{\small{ans}} in the first loop iteration\footnote{\textcolor{\rebuttalColor}{\reval only considers the first loop iteration for collecting program states, as reported in their artifact website~\cite{REvalArtifact}, failing to evaluate the ability of the models in reasoning through several loop iterations.}}. However, the model fails to correctly predict all the \texttt{\small{ans}}'s items in statements $9$ and $10$ (prompt2 response in blue and green). In response to prompt3 (next statement prediction) corresponding to statements $7$ (purple) and $9$ (blue), \gptf also fails to predict the ground truth value.}

\textcolor{\MRColor}{\reval does not implement any logic to combine the reasoning results of individual prompt responses to determine coherency of the runtime behavior for the entire program. Analysis of the responses corresponding to individual statements can also be confusing and not helpful to determine what resulted in an incorrect program execution and final output prediction: (1) prompt3 response corresponding to statement $7$ (purple) suggests that LLM terminates the loop after the first iteration (it predicts statement $8$ to execute next). However, prompt responses corresponding to statement $9$ (blue) indicate that LLM does iterate over the loop and populates \texttt{\small{ans}} with more items. (2) While statement $9$ is only supposed to append one element to the list if executed, prompt2 response for \texttt{\small{ans}} variable corresponding to statements $9$ (blue) and $10$ (green) are different in several indices. (3) Looking at the reasoning about statement $10$ (green), we can see that there is also a mismatch in predicting the value of \texttt{\small{ans}} and predicting output, although being the same variables. We will show how flow-sensitive prompting of \approach enables a more systematic reasoning analysis.}

\approach prompts LLMs to predict values of loop variables (\texttt{\small{e}} and \texttt{\small{o}}), loop iterable (\texttt{\small{zip(evens, odds)}}), predicate (\texttt{\small{len(evens)>len(odds)}}), branch (\texttt{\small{if}} statement), and output (\texttt{\small{ans}}). Figure~\ref{fig:illistrative-1}-c shows the response of \gptf to \approach prompt. \gptf simulates the execution from the program's start to the end, and predicted values, even if incorrect, \emph{coherently propagate throughout the program and result in incorrect output prediction}. \textcolor{\MRColor}{The divergence from the ground truth originates from the misprediction of the loop iterable sub-property, \texttt{\small{odds}}, at statement $6$, which also impacts the prediction of the compound loop variable \texttt{\small{zip(evens, odds)}} and the final output. That is, the LLM simulates program logic correctly after the simulation divergence point, but with the mispredicted value.} 
This analysis is achieved by prompting \gptf \emph{once}, compared to $10$ prompts in \reval that cannot provide helpful insights.

\begin{figure}
    \centering
    \includegraphics[width=0.85\linewidth]{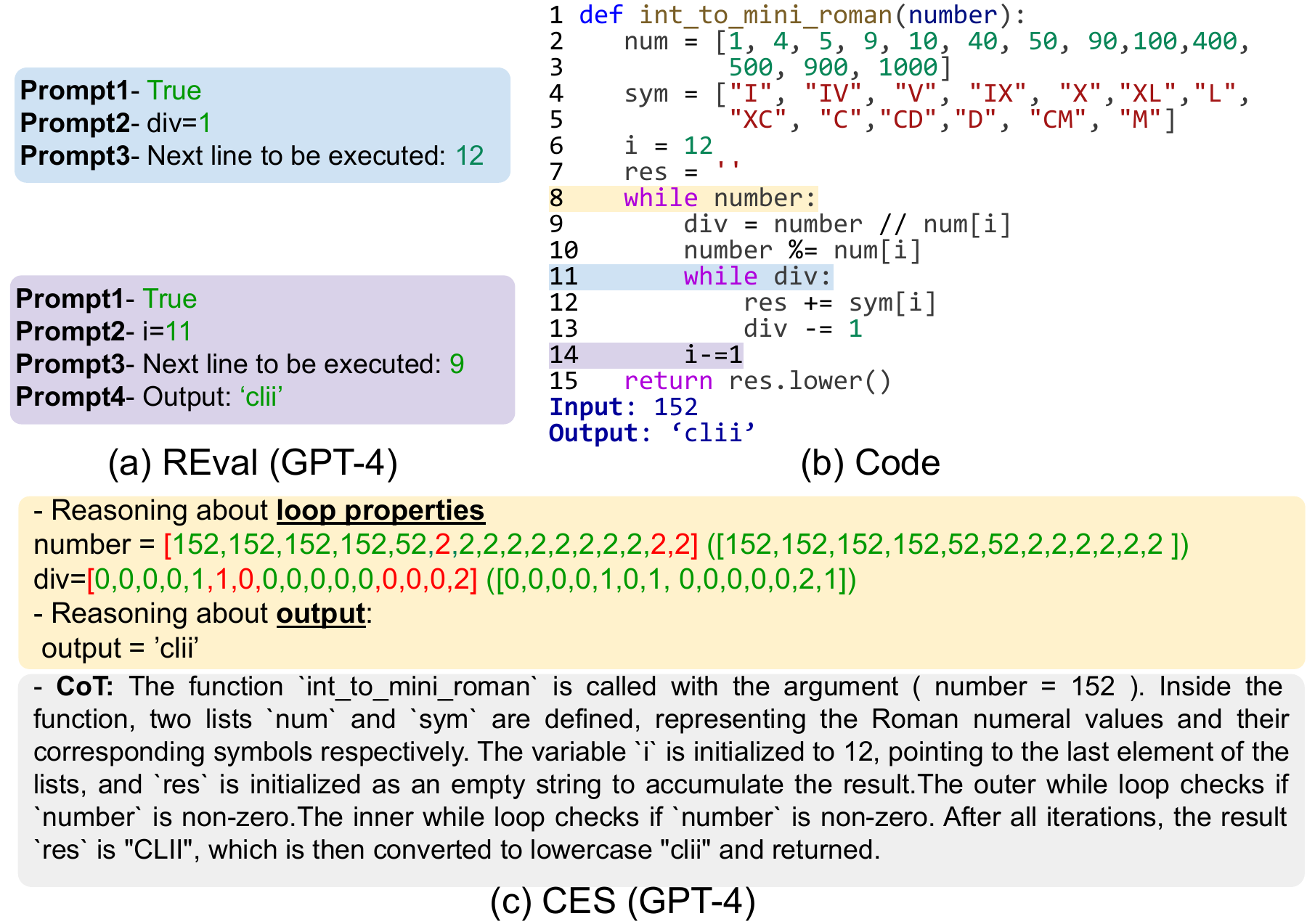}
    \vspace{-8pt}
    \caption{Output prediction of \gptf for HumanEval/156 (b); Reasoning analysis by \approach (c) and \reval (a)} 
    \vspace{-15pt}
    \label{fig:illistrative-example-2}
\end{figure}

\textcolor{\MRColor}{
Figure~\ref{fig:illistrative-example-2} demonstrates the ability of \approach in detecting suspiciously correct output prediction, compared to \reval and \codemind. All three approaches show that \gptf \emph{correctly} predicts the output. \reval even verdicts the reasoning is \emph{(incrementally) consistent}, since \gptf correctly predicts requested data in prompt1, prompt2, and prompt3 for its selected statements $11$ (blue) and $14$ (purple). On the other hand, \approach indicates that \gptf incorrectly predicts the value of loop variable \texttt{\small{numbers}} in statement $8$, starting from iteration $6$. This impacts the inner loop predictions, i.e., incorrect value prediction for \texttt{\small div} at the same iteration. In real execution, these incorrect predictions would propagate to an \emph{incorrect output prediction}. However, the predicted output from \gptf matches the ground truth. \approach marks this as \emph{incoherent} reasoning and discards the suspiciously correct output prediction. Further analysis shows that the \gptf Chain-of-Thought (CoT)~\cite{wei2022chain} only explains a high level of code without considering program states and variable values. Magically, \gptf changes the incorrect prediction through CoT in the response, matching the ground truth. This is likely due to data contamination, since \heval dataset is known to all LLMs. This example highlights the importance of eliminating suspiciously correct outputs to prevent inflated success in reasoning.}


\vspace{-5pt}
\section{Code Execution Simulation (\approach)}
\label{sec:approach}

A program is a set of statements for variable definition or assignment, introducing recursions to the logic (e.g., loops), introducing branches in the control flow (e.g., method calls, threading, conditional, and exception handling statements), or terminating the execution. \approach evaluates correctness of execution simulation in looping ($S_{loop}$), branching ($S_{branch}$), and returning ($S_{return}$) statements. The rationale is two-fold: (1) asking a model to predict all intermediate variable values can make the task complex, preventing proper responses and fair evaluation~\cite{liu2024lost}; (2) these statements identify the start or end of basic blocks, thereby capturing mispredictions in the assignment statements inside the block. We first formally define assessment properties (\S\ref{subsec:properties}) and then explain how \approach prompts models and evaluates these properties (\S\ref{subsec:metrics}). 

\vspace{-5pt}
\subsection{Program Properties}
\label{subsec:properties}

\textbf{Definition 1. Looping Properties.} A program may contain $m$ $(\ge0)$ loop statements, $l_j \in S_{loop}=\{l_1, \ldots, l_m\}$, with two main properties: loop variable ($V_{l_j}$) and loop iterable ($I_{l_j}$)\footnote{Note that loop iterable may not explicitly be specified in the loop statement, e.g., highlighted \texttt{\footnotesize while} statements in Figure~\ref{fig:illistrative-example-2}-b only have loop variables.}. The loop variable keeps track of the iterations, and the loop iterable defines the values or orders of the loop variable. \textcolor{\MRColor}{In the example of Figure~\ref{fig:illistrative-1}-b, the statement $6$ is a loop statement}. The loop variables are \texttt{\small e} and \texttt{\small o}, and the loop iterable is \texttt{\small zip(evens,odds)}. For each $l_j$ in a given program $P$ and concerning input(s) $I$, \approach prompts LLM to predict the values of all $V_{l_j}$s. The loop iterable can be a compound, i.e., consists of multiple variables or API calls. As a result, \approach asks the LLM to predict values of all sub-components. The rationale here is to correctly identify the root cause for simulation divergence. In the example of Figure~\ref{fig:illistrative-1}-b, the LLM may diverge from ground truth by mispredicting the values of \texttt{\small evens} or \texttt{\small odds}, or it may fail to understand the logic of \texttt{\small zip} API, mispredicting the return value of it even with correct values of \texttt{\small evens} and \texttt{\small odds}. 

\begin{figure}
    \centering
    \includegraphics[width=0.88\linewidth]{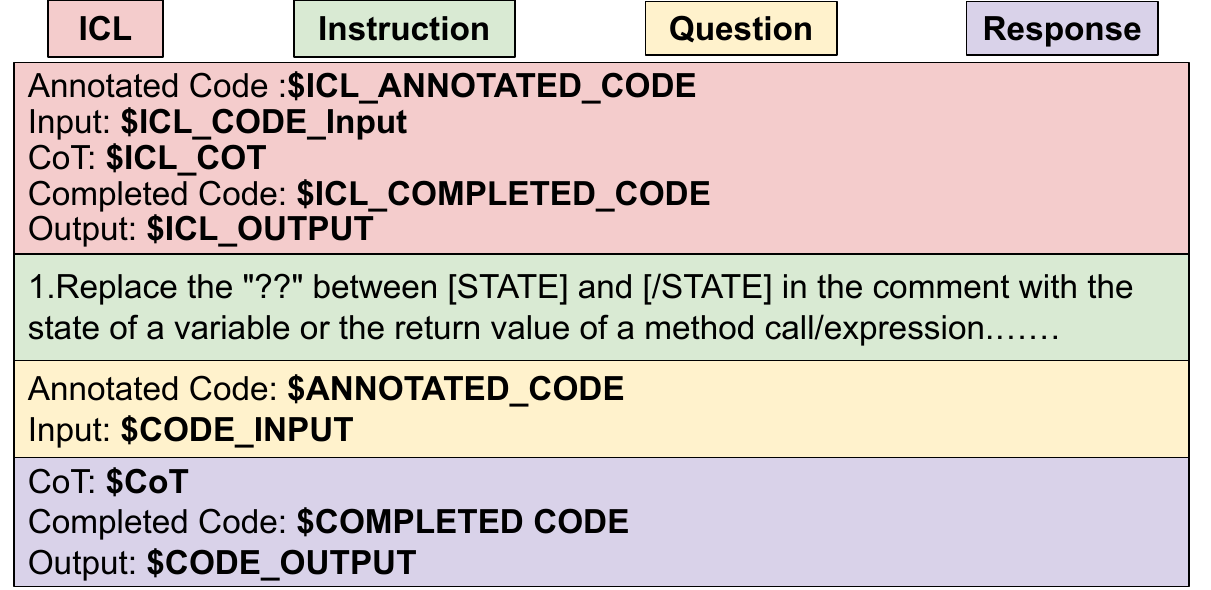}
    \vspace{-8pt}
    \caption{Prompt template used in \approach}
    \label{fig:prompt}
    \vspace{-10pt}
\end{figure}

\noindent\textbf{Definition 2. Branching Properties.} \approach currently only analyzes conditional branching statements ($S_{condition}$). A program may contain $n$ $(\ge0)$ conditional statements $c_j \in S_{condition}=\{c_1, \ldots, c_n\}$, each representing a branch $B_{c_j}$ in the control flow. A branch may be taken or not during execution, depending on the value of the conditional predicate $P_{c_j}$\footnote{For \texttt{\footnotesize{else}} statements, \approach only assesses $B_{c_j}$ as predicates do not exist.}. 
\textcolor{\MRColor}{In the example of Figure~\ref{fig:illistrative-1}-b, the result of conditional predicate \texttt{\small{len(evens)>len(odds)}} in statement $8$ determines whether the branch is taken.} A predicate can be a compound, i.e., consisting of multiple sub-predicates connected by logical operators. For example, a conditional statement \texttt{\small if(x>0 \&\& y<0)} consists of two sub-predicates, \texttt{\small x>0} and \texttt{\small y<0}. Compared to related research~\cite{chen2024reasoning, tufano2023predicting} that evaluate LLMs only for predicting the branches to be taken or not, \approach performs a finer-granularity assessment, asking LLMs to predict all sub-predicates, the whole predicate, and the branch. A misprediction at each could indicate a specific limitation of LLM in code reasoning: an LLM that correctly predicts sub-predicates but not the predicate may struggle with reasoning about complex logical expressions. An LLM that can correctly predict sub-predicates and predicate, but not the branch, struggles in understanding program construct semantics. As we will explain later in this section, this level of granularity also enables \approach to evaluate reasoning coherency (\S \ref{sec:valid-invalid}).  

\noindent\textbf{Definition 3. Return Properties.} A program may contain $k$ $(\ge0)$ return statements. A return statement $r_j \in S_{return}=\{r_1, \ldots, r_k\}$ defines the output of the program (a value to be returned or a message to be logged), $O_{r_j}$, once the execution terminates. \textcolor{\MRColor}{In Figure~\ref{fig:illistrative-1}-b, the program terminates at the return statement $10$.} Return statements can be compound, i.e., the program returns multiple variables at termination. In such cases, \approach breaks them down and evaluates the LLM in predicting individual values.

\vspace{-5pt}
\subsection{Prompting and Metrics}
\label{subsec:metrics}

\begin{figure}[t]
    \centering
    \includegraphics[width=0.85\linewidth]{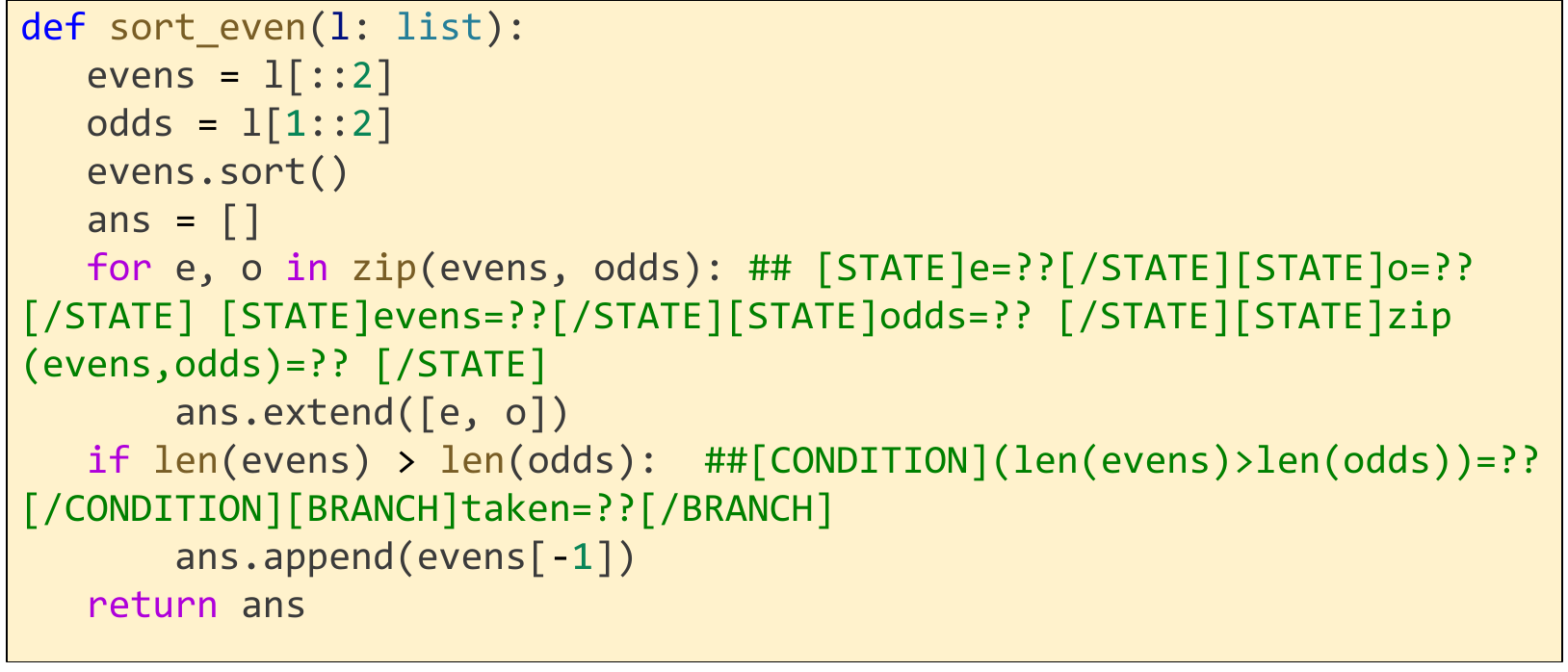}
    \vspace{-8pt}
    \caption{Annotated code in \approach's prompt}
    \label{fig:prompt-example}
    \vspace{-15pt}
\end{figure}

Figure~\ref{fig:prompt} shows the prompt template in \approach. Regardless of the number of loops, conditional, and return statements, and their nested level, \approach instructs the model to predict all properties during the execution simulation of given inputs. This enables a flow-sensitive assessment of execution reasoning and coherency analysis. \approach annotates all applicable locations in the code, which may or may not be covered during specific executions, preventing any hints to the model about the correct execution path. Figure~\ref{fig:prompt-example} demonstrates the annotated code in the prompt used to evaluate \gptf in the illustrative example of Figure~\ref{fig:illistrative-1}\footnote{\textcolor{\MRColor}{The complete prompt for the example can be found in \approach artifact website~\cite{website_prompt}}.}. \textcolor{\MRColor}{The annotations (prefixed with "\#\#") specify the properties to be predicted. LLMs should replace the placeholders marked as "??"  in the response with their predictions.}

\approach leverages in-context learning~\cite{brown2020language} to introduce code execution simulation task to LLMs. When in-context examples closely resemble the problem, performance improves significantly due to the model's generalization from familiar patterns~\cite{ye2023compositional,zhang2022active}. As a result, \approach constructs a pool of examples reflecting different combinations of programming constructs, including \textbf{if}, \textbf{elif}, \textbf{nested if}, \textbf{for loop}, \textbf{while loop}, \textbf{nested loop}, \textbf{if inside while loop}, \textbf{if outside while loop}, \textbf{if inside for loop}, \textbf{if outside for loop}, and \textbf{if inside nested loop}. When forming the prompt, it performs a lightweight flow-sensitive static analysis on $P$ to find the most relevant in-context example from the pool. It also prompts LLMs with implicit CoT. \textcolor{\rebuttalColor}{After receiving the response of model $M$ for simulating the execution of program $P$  under inputs $I$, \approach compares the ground truth and the predicted values for properties of  individual statements as below}: 

\vspace{-8pt}
\begin{small}
\begin{equation}
\label{eq:ces}
    CES(P,I,X_{y_j}) = \llbracket M(P,I,X_{y_{j}}) = GT(P,I,X_{y_{j}}) \rrbracket 
\end{equation}
\end{small}

\textcolor{\rebuttalColor}{
where $M(P,I,X_{y_{j}})$ and $GT(P,I,X_{y_{j}})$ denote the LLM prediction and ground truth values for $X_{y_j}$, respectively. $X_{y_j}$ can belong to $\{V_{i_j}\}_{j=0}^m$, $\{I_{l_j}\}_{j=0}^m$, $\{B_{c_j}\}_{j=0}^n$, $\{P_{c_j}\}_{j=0}^n$, or $\{O_{r_j}\}_{j=0}^k$. } The Iverson bracket $\llbracket$$\rrbracket$ returns $1$ if the condition in square brackets is satisfied and $0$ otherwise. $X_{y_j}$ can be compound, consisting of multiple sub-properties $X_{y_{j_w}}$. \approach automatically breaks all compound properties to sub-properties by parsing the code, and computes Equation~\ref{eq:ces} for each sub-property as well as the compound property. It uses the fine-grained evaluation of compound properties for reasoning coherency analysis (\S\ref{sec:valid-invalid}).

\section{Reasoning Coherency}
\label{sec:valid-invalid}

\approach considers execution simulations not compliant with common-sense execution logic as incoherent reasoning cases and marks anything else as coherent reasoning. It implements three coherency rules and \textcolor{\rebuttalColor}{checks the model's response for violations of} them.

\hspace{-4pt} $\bullet$ \hspace{1pt} \textbf{Rule 1.} Combining the sub-property predictions using code logic should match the compound property prediction. \textcolor{\rebuttalColor}{That is, for a compound property $X_{y_j}$ with $z$ sub-properties, if the program is executed with predicted sub-property values $X_{y_{j_w}} (1 \leq w \leq z)$ by LLM, the value of $X_{y_j}$ from execution should match that of prediction. \approach can automatically instrument the code with predicted sub-property values, generating an Instrumented Program (IP), executing that, and investigating the violation of Rule 1}:

\vspace{-10pt}
\begin{small}
\begin{equation}
\label{eq:invalid1}
\begin{split}
\exists \; j, X_{y_j} \in \left(\{I_{l_j}\}_{j=0}^m \cup \{P_{c_j}\}_{j=0}^n \cup \{O_{r_j}\}_{j=0}^k \right) \; \text{s.t.} \quad \quad \\
CES(P,I,X_{y_j}) \neq IP\bigl(CES(P,I,X_{y_{j_1}}), \ldots, CES(P,I,X_{y_{j_z}})\bigr)
\end{split}
\end{equation}
\end{small}
\vspace{-8pt}

Coherency concerning this rule does not imply correct prediction of all properties through simulation, but proper propagation of incorrect state. In Figure~\ref{fig:illistrative-1}-c, \gptf mispredicts values of \texttt{\small odds} variable. However, the predicted value of \texttt{\small zip(evens,odds)} properly combines the mispredicted value of \texttt{\small odds} and correctly predicted value of \texttt{\small evens}. Hence, the reasoning is coherent.

\hspace{-4pt} $\bullet$ \hspace{1pt} \textbf{Rule 2.} For a conditional statement, inconsistent conditional predicates and branch predictions demonstrate reasoning incoherence. This rule, again, is checked regardless of the ground truth values, since a mismatch between predicate values and branch decisions never happens in a real execution. \approach detects such cases as follows:

\vspace{-10pt}
\begin{small}
\begin{equation}
\label{eq:invalid2}
\underset{0 \leq j \leq n}{\exists} \; c_j 
\; \; \text{s.t.} \;\; CES(P,I,P_{c_j}) \; \neq \; CES(P,I,B_{c_j})
\end{equation}
\end{small}
\vspace{-10pt}

\hspace{-4pt} $\bullet$ \hspace{1pt} \textbf{Rule 3.} The most important case of reasoning incoherence is when the output is correctly predicted, but the prediction of \emph{at least} one of any other properties during simulation is incorrect. The illustrative example in Figure~\ref{fig:illistrative-example-2} shows such a case, which never happens in the real execution of the program, as the incorrect program state will be propagated to the output. As we will show (\S \ref{subsec:rq1}), LLMs are susceptible to reasoning incoherency due to hallucination, CoT shortcuts, and potential data leakage. \approach detects \textcolor{\rebuttalColor}{violation of Rule 3} using the following formula:

\vspace{-10pt}
\begin{small}
\begin{equation}
\label{eq:invalid3}
    \bigl(\underset{0 \leq j \leq k}{\exists} \; j \; \text{s.t.} \; CES(P,I,O_{r_j}) = 1\bigr) \; \land \; \Bigl(\prod_{f=1}^{m+n} CES(P,I,X_{y_f}) = 0\Bigr)
\end{equation}
\end{small}
\vspace{-8pt}

\textcolor{\rebuttalColor}{where $X_{y_f}$ belongs to $\{V_{i_f}\}_{f=0}^m \cup \{I_{l_f}\}_{f=0}^m \cup \{B_{c_f}\}_{f=0}^n \cup \{P_{c_f}\}_{f=0}^n$ statements that appear \emph{before} $O_{r_j}$.} For compound properties, Formula~\ref{eq:invalid3} only considers the incorrect prediction of the property, regardless of sub-property predictions. This is because a misprediction at a coarser grain is more likely to propagate to the rest of the simulation. Any incoherence between sub-properties and compound property will be captured by Rule 1. 

If any of the above cases happen, \approach marks the execution simulation reasoning as incoherent. The incoherent reasoning cases, even if the output prediction is correct, will be discarded and won't be considered as LLM success. 
\vspace{-8pt}
\subsection{Diagnosis}
\label{sec:approach-diagnostic}

\sloppy \approach offers two automated diagnostic analyses: identifying where the simulation diverges from ground truth (coherent reasoning followed by incorrect output prediction) and detecting suspiciously correct output predictions (correct output prediction achieved through incoherent reasoning).

When the reasoning is \emph{coherent} and the output prediction is \emph{incorrect}, \approach performs a forward flow-sensitive static analysis on the program $P$ and selects the first statement $j$ with $CES(M,P,I,X_{y_j}) = 0$. Given that the reasoning process has been checked for coherence firsthand, the first incorrect prediction likely propagates in the execution simulation, causing the LLM to mispredict subsequent properties. In the illustrative example of Figure~\ref{fig:illistrative-1}, \approach automatically detects the misprediction of \texttt{\small{e}} as a simulation divergence point. 

\approach can \textcolor{\CRColor}{efficiently} identify suspiciously correct outputs upon receiving the response from the model. To that end, it checks for violations of coherence rules to identify incoherent reasoning cases. Violation of Rule 3 (Formula~\ref{eq:invalid3}) directly indicates suspiciously correct output prediction. Violations of the other two rules (Formulas~\ref{eq:invalid1}--\ref{eq:invalid2}), if accompanied by correct output prediction, are marked as suspiciously correct cases.

\vspace{-10pt}
\section{Reasoning Consistency}
\label{approach:strong-weak-random}

LLM-generated code should be comprehensively tested across all the execution paths in the program. Otherwise, one can falsely claim an incorrect code to be correct. Similarly, an LLM cannot claim victory on code execution reasoning until it can simulate all execution paths correctly. \approach introduces the spectrum of reasoning consistency, considering the model's \approach performance across different execution paths. 

\begin{figure}
    \centering
    \includegraphics[width=0.90\linewidth]{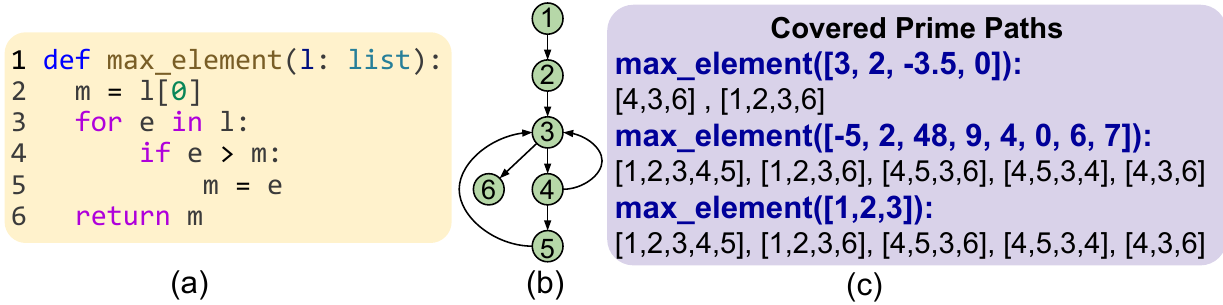}
    \vspace{-8pt}
    \caption{HumanEval/35 program (a), its control flow graph (b), and the prime path coverage of three tests (c)}
    \vspace{-15pt}
    \label{fig:enter-label}
\end{figure}

Figure~\ref{fig:enter-label} shows a program with its control flow graph (CFG). 
The loop in this simple program introduces an unbounded number of execution paths. Due to the prevalence of loops and recursion in programs, \approach considers \emph{prime paths} in the CFG to bound the number of execution paths. A prime path in a cyclic CFG is a maximal path between two arbitrary nodes that does not visit any node more than once, except for the start and end~\cite{ammann2017introduction}. \textcolor{\rebuttalColor}{Prime paths not only bound the number of execution paths, but cover all CFG edges and execution sequences, enabling completeness of \approach consistency analysis. The CFG in Figure~\ref{fig:enter-label}-b has \emph{five} prime paths: $[1,2,3,4,5]$, $[1,2,3,6]$, $[4,5,3,6]$, $[4,5,3,4]$, and $[4,3,6]$.} Figure~\ref{fig:enter-label}-c shows three tests for the program of Figure~\ref{fig:enter-label}-a and prime paths they cover during test execution. 

Given a program $P$, \textcolor{\rebuttalColor}{tests $\{t_1, \ldots, t_g\}$}, and prime paths covered by tests $\{cov_1, \ldots cov_g\}$, \approach defines the spectrum of code reasoning consistency as follows: 

\textbf{Definition 4. Weak Consistency ($WC$).} An LLM achieves a weak level of code execution reasoning consistency on program $P$ if it can correctly simulate the execution of $P$ under all tests covering \emph{same} prime paths, but may fail in simulating a test covering different prime paths. \approach checks weak consistency as:

\vspace{-7pt}
\begin{footnotesize}
\begin{equation} 
\label{eq:weakr}
    \begin{split}
    WC \mathrel{:=} [\underset{1 \leq z,q \leq g}{\exists} z \neq q  \; \; cov_z = cov_q  \; \land \quad \quad \quad \quad \quad \quad \quad \quad \quad \\ 
    \forall \; z \neq q \; \; \bigl( cov_z = cov_q \bigr) \Rightarrow \bigl( CES(P,i_z,X_{y_j}) = CES(P,i_q,X_{y_j}) = 1 \bigr)]
    \end{split}
\end{equation}
\end{footnotesize}

\textcolor{\rebuttalColor}{The predicate $cov_z=cov_q$ will be \texttt{\small{True}} if $t_z$ and $t_q$ cover the same number of identical prime paths. In Figure~\ref{fig:enter-label}-c, the last two tests cover the same prime paths and have the same coverage. Although the prime paths covered by the first test are a subset of those covered by other tests, \approach considers them to have different coverage. 
$WC$ will be \texttt{\small{True}} if there is at least a pair of tests covering the same prime paths (first conjunct), and LLM successfully simulates execution of all tests covering the same prime paths (second conjunct). The second conjunct will be \texttt{\small{False}} if LLM incorrectly simulates two tests covering the same prime paths, or if it correctly simulates execution of two tests covering different prime paths.}

\vspace{3pt}
\textbf{Definition 5. Strong Consistency ($SC$).} An LLM achieves a strong level of code execution reasoning consistency on program $P$, if it can correctly simulate the execution of $P$ under all tests covering \emph{different} prime paths, and if exist, \emph{same} prime paths.   

\vspace{-7pt}
\begin{footnotesize}
\begin{equation} 
\label{eq:strongr}
    \begin{split}
    SC \mathrel{:=} [\underset{1 \leq z,q \leq g}{\exists} z \neq q \; \; cov_z \neq cov_q \; \land \quad \quad \quad \\ 
    \forall \; z \neq q \; \; CES(P,i_z,X_{y_j}) = CES(P,i_q,X_{y_j}) = 1]
    \end{split}
\end{equation}
\end{footnotesize}

\textcolor{\rebuttalColor}{$SC$ will be \texttt{\small{True}} if there exist at least two tests covering different prime paths (first conjunct), and LLM correctly simulates all the tests, regardless of the prime paths they cover (second conjunct).}

LLM's reasoning consistency is \emph{random} on program $P$, otherwise. 
\textcolor{\rebuttalColor}{If there are no tests with different covered prime paths in the evaluation of an LLM, and it correctly simulates all (the otherwise clause in Formula~\ref{eq:weakr}), \approach considers it as a weak consistency. The only exception is when all the prime paths are covered by each test, where \approach considers it as strong consistency.} In the example of Figure~\ref{fig:enter-label}, \gptf consistently and correctly simulates the program execution across three tests with similar or different prime paths covered, achieving strong consistency. Strong consistency does not require test inputs covering all the prime paths, but rather a combination of similar (under different inputs) and different coverages. Even with such slack, we observe that LLMs achieve strong consistency only for a handful of programs in the \heval benchmark (\S\ref{rq:reasoning-consistency}).

\begin{table*}[t]
\caption{Performance of LLMs in \approach considering all the three sampled tests per each HumanEval program. \textbf{CO}: programs with conditional statements only, \textbf{LO}: programs with loops only, and \textbf{LC}: programs with both loops and conditional statements. We highlight the top three best-performing models with \textcolor{red}{red ($1^{st}$)}, \textcolor{teal}{green ($2^{nd}$)}, and \textcolor{blue}{blue ($3^{rd}$)}.}
\vspace{-8pt}
\label{table:rq1}
\setlength{\tabcolsep}{2pt}
\resizebox{0.98\textwidth}{!}{
\begin{tabular}{|c|cc|cccccccccccccccc|}
\hline
\multirow{2}{*}{\textbf{Programs}} &
\multicolumn{2}{|c|}{\multirow{2}{*}{\textbf{Predictions}}} &
\multicolumn{16}{c|}{\textbf{Subject LLMs}} 
\\ \cline{4-19} 
& 
\multicolumn{2}{|c|}{} &
\multicolumn{5}{c|}{\textbf{CodeLlama}}  & 
\multicolumn{3}{c|}{\textbf{DeepSeekCoder}}  &
\multicolumn{1}{c|}{\textbf{\begin{tabular}[c]{@{}c@{}}MagiCoder-S\end{tabular}}} &
\multicolumn{1}{c|}{\textbf{\begin{tabular}[c]{@{}c@{}}SemCoder-S\end{tabular}}} &
\multicolumn{1}{c|}{\textbf{\begin{tabular}[c]{@{}c@{}}StarCoder2\end{tabular}}} &
\multicolumn{1}{c|}{\textbf{\begin{tabular}[c]{@{}c@{}}Gemini-1.5\end{tabular}}} &
\multicolumn{1}{c|}{\textbf{\begin{tabular}[c]{@{}c@{}}GPT-4\end{tabular}}} &
\multicolumn{1}{c|}{\textbf{\begin{tabular}[c]{@{}c@{}}\textcolor{\rebuttalColor}{DeepSeek-}\end{tabular}}} &
\multicolumn{1}{c|}{\textbf{\begin{tabular}[c]{@{}c@{}}\textcolor{\rebuttalColor}{Gemini-2.5}\end{tabular}}} &
\multicolumn{1}{c|}{\textbf{\begin{tabular}[c]{@{}c@{}}\textcolor{\rebuttalColor}{o4-mini}\end{tabular}}} \\

& 
\multicolumn{2}{|c|}{} &
\multicolumn{1}{c|}{\textbf{\begin{tabular}[c]{@{}c@{}}(Inst-7b)\end{tabular}}} &
\multicolumn{1}{c|}{\textbf{\begin{tabular}[c]{@{}c@{}}(Base-7b)\end{tabular}}} &
\multicolumn{1}{c|}{\textbf{\begin{tabular}[c]{@{}c@{}}(Inst-13b)\end{tabular}}} &
\multicolumn{1}{c|}{\textbf{\begin{tabular}[c]{@{}c@{}}(Base-13b)\end{tabular}}} &
\multicolumn{1}{c|}{\textbf{\begin{tabular}[c]{@{}c@{}}(Inst-34b)\end{tabular}}} &
\multicolumn{1}{c|}{\textbf{\begin{tabular}[c]{@{}c@{}}(Inst-6.7b)\end{tabular}}} &
\multicolumn{1}{c|}{\textbf{\begin{tabular}[c]{@{}c@{}}(Base-6.7b)\end{tabular}}} &
\multicolumn{1}{c|}{\textbf{\begin{tabular}[c]{@{}c@{}}(Inst-33b)\end{tabular}}} &
\multicolumn{1}{c|}{\textbf{\begin{tabular}[c]{@{}c@{}}(6.7b)\end{tabular}}} &
\multicolumn{1}{c|}{\textbf{\begin{tabular}[c]{@{}c@{}}(6.7b)\end{tabular}}} &
\multicolumn{1}{c|}{\textbf{\begin{tabular}[c]{@{}c@{}}(15b)\end{tabular}}} &
\multicolumn{1}{c|}{\textbf{\begin{tabular}[c]{@{}c@{}}(Pro)\end{tabular}}} &
\multicolumn{1}{c|}{\textbf{\begin{tabular}[c]{@{}c@{}}\end{tabular}}} &
\multicolumn{1}{c|}{\textbf{\begin{tabular}[c]{@{}c@{}}R1\end{tabular}}} &
\multicolumn{1}{c|}{\textbf{\begin{tabular}[c]{@{}c@{}}\textcolor{\rebuttalColor}{(Pro)}\end{tabular}}} &
\multicolumn{1}{c|}{\textbf{\begin{tabular}[c]{@{}c@{}}\end{tabular}}}\\ 
\hline

\multirow{4}{*}{\textbf{CO (24)}} & 
\multicolumn{1}{|c|}{\multirow{2}{*}{\cellcolor{correctres}\textbf{\begin{tabular}[c]{@{}c@{}}Coherent \\ Reasoning \end{tabular}}}} &
\cellcolor{correctout}\textbf{\begin{tabular}[c]{@{}c@{}}Correct Output\end{tabular}} &
\multicolumn{1}{c|}{23.37\%} &
\multicolumn{1}{c|}{20.90\%} &
\multicolumn{1}{c|}{25.97\%} &
\multicolumn{1}{c|}{23.38\%} &
\multicolumn{1}{c|}{30.88\%} &
\multicolumn{1}{c|}{38.89\%} &
\multicolumn{1}{c|}{37.50\%} &
\multicolumn{1}{c|}{41.67\%} &
\multicolumn{1}{c|}{29.17\%} &
\multicolumn{1}{c|}{{47.22\%}} &
\multicolumn{1}{c|}{40.28\%} &
\multicolumn{1}{c|}{{54.17\%}} &
\multicolumn{1}{c|}{\textcolor{teal}{81.94\%}} &
\multicolumn{1}{c|}{73.61\%} &
\multicolumn{1}{c|}{\textcolor{red}{90.28\%}} &
\multicolumn{1}{c|}{\textcolor{blue}{80.56\%}}
\\ \cline{3-19} 

& 
\multicolumn{1}{|c|}{\cellcolor{correctres}\textbf{Reasoning}} &
  \cellcolor{incorrectout}\textbf{\begin{tabular}[c]{@{}c@{}}Incorrect Output\end{tabular}} &
\multicolumn{1}{c|}{42.86\%} &
\multicolumn{1}{c|}{53.64\%} &
\multicolumn{1}{c|}{44.16\%} &
\multicolumn{1}{c|}{45.45\%} &
\multicolumn{1}{c|}{50.00\%} &
\multicolumn{1}{c|}{36.11\%} &
\multicolumn{1}{c|}{51.39\%} &
\multicolumn{1}{c|}{36.11\%} &
\multicolumn{1}{c|}{52.78\%} &
\multicolumn{1}{c|}{38.89\%} &
\multicolumn{1}{c|}{50.00\%} &
\multicolumn{1}{c|}{30.56\%} &
\multicolumn{1}{c|}{8.33\%} &
\multicolumn{1}{c|}{9.72\%} &
\multicolumn{1}{c|}{2.78\%} &
\multicolumn{1}{c|}{0.00\%} 
  \\ \cline{3-19} 
 
& 
  \multicolumn{1}{|c|}{\multirow{2}{*}{\cellcolor{incorrectres}\textbf{\begin{tabular}[c]{@{}c@{}}Incoherent \\ Reasoning \end{tabular}}}} &
  \cellcolor{correctout}\textbf{\begin{tabular}[c]{@{}c@{}}Correct Output\end{tabular}} &
\multicolumn{1}{c|}{16.88\%} &
\multicolumn{1}{c|}{25.47\%} &
\multicolumn{1}{c|}{22.08\%} &
\multicolumn{1}{c|}{24.68\%} &
\multicolumn{1}{c|}{19.12\%} &
\multicolumn{1}{c|}{19.44\%} &
\multicolumn{1}{c|}{11.11\%} &
\multicolumn{1}{c|}{20.83\%} &
\multicolumn{1}{c|}{18.06\%} &
\multicolumn{1}{c|}{12.50\%} &
\multicolumn{1}{c|}{6.94\%} &
\multicolumn{1}{c|}{13.89\%} &
\multicolumn{1}{c|}{9.72\%} & 
\multicolumn{1}{c|}{16.67\%} &
\multicolumn{1}{c|}{6.94\%} &
\multicolumn{1}{c|}{19.44\%} 
  \\ \cline{3-19} 
 
& 
  \multicolumn{1}{|c|}{\cellcolor{incorrectres}\textbf{Reasoning}} 
  &\cellcolor{incorrectout}\textbf{\begin{tabular}[c]{@{}c@{}}Incorrect Output\end{tabular}} 
  &\multicolumn{1}{c|}{16.88\%} 
  &\multicolumn{1}{c|}{0.00\%} 
  &\multicolumn{1}{c|}{7.79\%} 
  &\multicolumn{1}{c|}{6.49\%} 
  &\multicolumn{1}{c|}{0.00\%}
  &\multicolumn{1}{c|}{0.00\%} 
  &\multicolumn{1}{c|}{0.00\%}
  &\multicolumn{1}{c|}{1.39\%} 
  &\multicolumn{1}{c|}{0.00\%}
  &\multicolumn{1}{c|}{1.39\%}
  &\multicolumn{1}{c|}{2.78\%}
  &\multicolumn{1}{c|}{1.39\%}
  &\multicolumn{1}{c|}{0.00\%}
  &\multicolumn{1}{c|}{0.00\%}
  &\multicolumn{1}{c|}{0.00\%}
  &\multicolumn{1}{c|}{0.00\%}
  \\ \hline

\multirow{4}{*}{\textbf{LO (12)}} & 
\multicolumn{1}{|c|}{\multirow{2}{*}{\cellcolor{correctres}\textbf{\begin{tabular}[c]{@{}c@{}}Coherent \\ Reasoning \end{tabular}}}} &
\cellcolor{correctout}\textbf{\begin{tabular}[c]{@{}c@{}}Correct Output\end{tabular}} &
\multicolumn{1}{c|}{8.11\%} &
\multicolumn{1}{c|}{0.00\%} &
\multicolumn{1}{c|}{40.54\%} &
\multicolumn{1}{c|}{35.14\%} &
\multicolumn{1}{c|}{35.29\%} &
\multicolumn{1}{c|}{{47.22\%}} &
\multicolumn{1}{c|}{38.89\%} &
\multicolumn{1}{c|}{{52.78\%}} &
\multicolumn{1}{c|}{41.67\%} &
\multicolumn{1}{c|}{41.67\%} &
\multicolumn{1}{c|}{40.54\%} &
\multicolumn{1}{c|}{38.89\%} &
\multicolumn{1}{c|}{55.56\%} &
\multicolumn{1}{c|}{\textcolor{red}{77.78\%}} &
\multicolumn{1}{c|}{\textcolor{blue}{75.00\%}} &
\multicolumn{1}{c|}{\textcolor{red}{77.78\%}}
\\ \cline{3-19} 

& 
\multicolumn{1}{|c|}{\cellcolor{correctres}\textbf{Reasoning}} &
  \cellcolor{incorrectout}\textbf{\begin{tabular}[c]{@{}c@{}}Incorrect Output\end{tabular}} &
\multicolumn{1}{c|}{64.86\%} &
\multicolumn{1}{c|}{62.16\%} &
\multicolumn{1}{c|}{43.24\%} &
\multicolumn{1}{c|}{51.35\%} &
\multicolumn{1}{c|}{61.76\%} &
\multicolumn{1}{c|}{36.11\%} &
\multicolumn{1}{c|}{52.78\%} &
\multicolumn{1}{c|}{30.56\%} &
\multicolumn{1}{c|}{44.44\%} &
\multicolumn{1}{c|}{47.22\%} &
\multicolumn{1}{c|}{52.78\%} &
\multicolumn{1}{c|}{36.11\%} &
\multicolumn{1}{c|}{5.56\%} & 
\multicolumn{1}{c|}{19.44\%} &
\multicolumn{1}{c|}{0.00\%} &
\multicolumn{1}{c|}{0.00\%}
  \\ \cline{3-19} 
 
& 
  \multicolumn{1}{|c|}{\multirow{2}{*}{\cellcolor{incorrectres}\textbf{\begin{tabular}[c]{@{}c@{}}Incoherent \\ Reasoning \end{tabular}}}} &
  \cellcolor{correctout}\textbf{\begin{tabular}[c]{@{}c@{}}Correct Output\end{tabular}} &
\multicolumn{1}{c|}{27.03\%} &
\multicolumn{1}{c|}{37.84\%} &
\multicolumn{1}{c|}{16.22\%} &
\multicolumn{1}{c|}{13.51\%} &
\multicolumn{1}{c|}{22.55\%} &
\multicolumn{1}{c|}{16.67\%} &
\multicolumn{1}{c|}{8.33\%} &
\multicolumn{1}{c|}{16.67\%} &
\multicolumn{1}{c|}{13.89\%} &
\multicolumn{1}{c|}{11.11\%} &
\multicolumn{1}{c|}{8.33\%} &
\multicolumn{1}{c|}{27.78\%} &
\multicolumn{1}{c|}{38.89\%} & 
\multicolumn{1}{c|}{2.78\%} &
\multicolumn{1}{c|}{25.00\%} &
\multicolumn{1}{c|}{22.22\%} 
  \\ \cline{3-19} 
 
& 
  \multicolumn{1}{|c|}{\cellcolor{incorrectres}\textbf{Reasoning}} 
  &\cellcolor{incorrectout}\textbf{\begin{tabular}[c]{@{}c@{}}Incorrect Output\end{tabular}} 
  &\multicolumn{1}{c|}{0.00\%} 
  &\multicolumn{1}{c|}{0.00\%} 
  &\multicolumn{1}{c|}{0.00\%} 
  &\multicolumn{1}{c|}{0.00\%} 
  &\multicolumn{1}{c|}{0.00\%}
  &\multicolumn{1}{c|}{0.00\%} 
  &\multicolumn{1}{c|}{0.00\%}
  &\multicolumn{1}{c|}{0.00\%} 
  &\multicolumn{1}{c|}{0.00\%}
  &\multicolumn{1}{c|}{0.00\%}
  &\multicolumn{1}{c|}{0.00\%}
  &\multicolumn{1}{c|}{0.00\%}
  &\multicolumn{1}{c|}{0.00\%}
  &\multicolumn{1}{c|}{0.00\%}
  &\multicolumn{1}{c|}{0.00\%}
  &\multicolumn{1}{c|}{0.00\%}  
  \\ \hline

\multirow{4}{*}{\textbf{LC (75)}} & 
\multicolumn{1}{|c|}{\multirow{2}{*}{\cellcolor{correctres}\textbf{\begin{tabular}[c]{@{}c@{}}Coherent \\ Reasoning \end{tabular}}}} &
\cellcolor{correctout}\textbf{\begin{tabular}[c]{@{}c@{}}Correct Output\end{tabular}} &
\multicolumn{1}{c|}{3.69\%} &
\multicolumn{1}{c|}{2.46\%} &
\multicolumn{1}{c|}{7.37\%} &
\multicolumn{1}{c|}{4.15\%} &
\multicolumn{1}{c|}{6.86\%} &
\multicolumn{1}{c|}{11.56\%} &
\multicolumn{1}{c|}{7.11\%} &
\multicolumn{1}{c|}{{15.56\%}} &
\multicolumn{1}{c|}{11.56\%} &
\multicolumn{1}{c|}{10.22\%} &
\multicolumn{1}{c|}{{16.00\%}} &
\multicolumn{1}{c|}{12.44\%} &
\multicolumn{1}{c|}{36.00\%} &
\multicolumn{1}{c|}{\textcolor{blue}{48.44\%}} &
\multicolumn{1}{c|}{\textcolor{teal}{48.88\%}} &
\multicolumn{1}{c|}{\textcolor{red}{60.44\%}}
\\ \cline{3-19} 

& 
\multicolumn{1}{|c|}{\cellcolor{correctres}\textbf{Reasoning}} &
  \cellcolor{incorrectout}\textbf{\begin{tabular}[c]{@{}c@{}}Incorrect Output\end{tabular}} &
\multicolumn{1}{c|}{68.66\%} &
\multicolumn{1}{c|}{77.72\%} &
\multicolumn{1}{c|}{71.43\%} &
\multicolumn{1}{c|}{64.98\%} &
\multicolumn{1}{c|}{70.59\%} &
\multicolumn{1}{c|}{56.89\%} &
\multicolumn{1}{c|}{68.89\%} &
\multicolumn{1}{c|}{55.56\%} &
\multicolumn{1}{c|}{65.33\%} &
\multicolumn{1}{c|}{61.33\%} &
\multicolumn{1}{c|}{53.33\%} &
\multicolumn{1}{c|}{43.56\%} &
\multicolumn{1}{c|}{18.67\%} &
\multicolumn{1}{c|}{6.22\%} &
\multicolumn{1}{c|}{3.59\%} &
\multicolumn{1}{c|}{3.56\%}
  \\ \cline{3-19} 
 
& 
  \multicolumn{1}{|c|}{\multirow{2}{*}{\cellcolor{incorrectres}\textbf{\begin{tabular}[c]{@{}c@{}}Incoherent \\ Reasoning \end{tabular}}}} &
  \cellcolor{correctout}\textbf{\begin{tabular}[c]{@{}c@{}}Correct Output\end{tabular}} &
\multicolumn{1}{c|}{26.73\%} &
\multicolumn{1}{c|}{19.82\%} &
\multicolumn{1}{c|}{21.20\%} &
\multicolumn{1}{c|}{30.41\%} &
\multicolumn{1}{c|}{22.55\%} &
\multicolumn{1}{c|}{30.67\%} &
\multicolumn{1}{c|}{24.00\%} &
\multicolumn{1}{c|}{28.44\%} &
\multicolumn{1}{c|}{22.67\%} &
\multicolumn{1}{c|}{28.00\%} &
\multicolumn{1}{c|}{30.22\%} &
\multicolumn{1}{c|}{40.44\%} &
\multicolumn{1}{c|}{45.33\%} &
\multicolumn{1}{c|}{45.33\%} &
\multicolumn{1}{c|}{47.53\%} &
\multicolumn{1}{c|}{36.00\%}
  \\ \cline{3-19} 
 
& 
  \multicolumn{1}{|c|}{\cellcolor{incorrectres}\textbf{Reasoning}} 
  &\cellcolor{incorrectout}\textbf{\begin{tabular}[c]{@{}c@{}}Incorrect Output\end{tabular}} 
  &\multicolumn{1}{c|}{0.88\%} 
  &\multicolumn{1}{c|}{0.00\%} 
  &\multicolumn{1}{c|}{0.00\%} 
  &\multicolumn{1}{c|}{0.44\%} 
  &\multicolumn{1}{c|}{0.00\%}
  &\multicolumn{1}{c|}{0.89\%} 
  &\multicolumn{1}{c|}{0.00\%}
  &\multicolumn{1}{c|}{0.44\%} 
  &\multicolumn{1}{c|}{0.44\%}
  &\multicolumn{1}{c|}{0.44\%}
  &\multicolumn{1}{c|}{0.44\%}
  &\multicolumn{1}{c|}{1.78\%}
  &\multicolumn{1}{c|}{0.00\%}
  &\multicolumn{1}{c|}{0.00\%}
  &\multicolumn{1}{c|}{0.00\%}
  &\multicolumn{1}{c|}{0.00\%}
  \\ \hline

\multirow{2}{*}{\textbf{\begin{tabular}[c]{@{}c@{}}Others (53)\end{tabular}}} &
  \multicolumn{2}{|c|}{\cellcolor{correctout}\textbf{\begin{tabular}[c]{@{}c@{}}Correct Output\end{tabular}}} &
\multicolumn{1}{c|}{40.37\%} &
\multicolumn{1}{c|}{37.93\%} &
\multicolumn{1}{c|}{42.59\%} &
\multicolumn{1}{c|}{34.19\%} &
\multicolumn{1}{c|}{46.15\%} &
\multicolumn{1}{c|}{52.83\%} &
\multicolumn{1}{c|}{50.94\%} &
\multicolumn{1}{c|}{{69.18\%}} &
\multicolumn{1}{c|}{62.26\%} &
\multicolumn{1}{c|}{62.89\%} &
\multicolumn{1}{c|}{{67.30\%}} &
\multicolumn{1}{c|}{65.41\%} &
\multicolumn{1}{c|}{\textcolor{blue}{88.05\%}} &
\multicolumn{1}{c|}{\textcolor{teal}{90.57\%}} &
\multicolumn{1}{c|}{87.42\%} &
\multicolumn{1}{c|}{\textcolor{red}{91.19\%}}
\\ \cline{2-19} 
 &
  \multicolumn{2}{|c|}
  {\cellcolor{incorrectout}\textbf{\begin{tabular}[c]{@{}c@{}}Incorrect Output\end{tabular}}} &
\multicolumn{1}{c|}{59.63\%} &
\multicolumn{1}{c|}{62.07\%} &
\multicolumn{1}{c|}{57.41\%} &
\multicolumn{1}{c|}{65.81\%} &
\multicolumn{1}{c|}{53.85\%} &
\multicolumn{1}{c|}{47.17\%} &
\multicolumn{1}{c|}{49.06\%} &
\multicolumn{1}{c|}{30.82\%} &
\multicolumn{1}{c|}{37.74\%} &
\multicolumn{1}{c|}{37.11\%} &
\multicolumn{1}{c|}{32.70\%} &
\multicolumn{1}{c|}{34.59\%} &
\multicolumn{1}{c|}{11.95\%} &
\multicolumn{1}{c|}{9.43\%} &
\multicolumn{1}{c|}{12.58\%} &
\multicolumn{1}{c|}{8.81\%}
\\ \hline

\multirow{4}{*}{\textbf{Total (164)}} & 
\multicolumn{1}{|c|}{\multirow{2}{*}{\cellcolor{correctres}\textbf{\begin{tabular}[c]{@{}c@{}}Coherent \\ Reasoning \end{tabular}}}} &
\cellcolor{correctout}\textbf{\begin{tabular}[c]{@{}c@{}}Correct Output\end{tabular}} &
\multicolumn{1}{c|}{18.40\%} &
\multicolumn{1}{c|}{16.44\%} &
\multicolumn{1}{c|}{19.14\%} &
\multicolumn{1}{c|}{17.49\%} &
\multicolumn{1}{c|}{25.17\%} &
\multicolumn{1}{c|}{31.51\%} &
\multicolumn{1}{c|}{28.04\%} &
\multicolumn{1}{c|}{{39.43\%}} &
\multicolumn{1}{c|}{32.72\%} &
\multicolumn{1}{c|}{34.96\%} &
\multicolumn{1}{c|}{37.80\%} &
\multicolumn{1}{c|}{{38.21\%}} &
\multicolumn{1}{c|}{60.98\%} &
\multicolumn{1}{c|}{\textcolor{blue}{67.61\%}} &
\multicolumn{1}{c|}{\textcolor{teal}{69.08\%}} &
\multicolumn{1}{c|}{\textcolor{red}{74.39\%}}
\\ \cline{3-19} 

& 
\multicolumn{1}{|c|}{\cellcolor{correctres}\textbf{Reasoning}} &
  \cellcolor{incorrectout}\textbf{\begin{tabular}[c]{@{}c@{}}Incorrect Output\end{tabular}} &
\multicolumn{1}{c|}{61.85\%} &
\multicolumn{1}{c|}{66.69\%} &
\multicolumn{1}{c|}{61.11\%} &
\multicolumn{1}{c|}{67.07\%} &
\multicolumn{1}{c|}{61.47\%} &
\multicolumn{1}{c|}{49.28\%} &
\multicolumn{1}{c|}{58.75\%} &
\multicolumn{1}{c|}{42.88\%} &
\multicolumn{1}{c|}{53.05\%} &
\multicolumn{1}{c|}{49.39\%} &
\multicolumn{1}{c|}{46.14\%} &
\multicolumn{1}{c|}{38.22\%} &
\multicolumn{1}{c|}{14.02\%} &
\multicolumn{1}{c|}{7.50\%} &
\multicolumn{1}{c|}{7.88\%} &
\multicolumn{1}{c|}{6.06\%}
  \\ \cline{3-19} 
 
& 
  \multicolumn{1}{|c|}{\multirow{2}{*}{\cellcolor{incorrectres}\textbf{\begin{tabular}[c]{@{}c@{}}Incoherent \\ Reasoning \end{tabular}}}} &
  \cellcolor{correctout}\textbf{\begin{tabular}[c]{@{}c@{}}Correct Output\end{tabular}} &
 \multicolumn{1}{c|}{16.67\%} &
\multicolumn{1}{c|}{16.87\%} &
\multicolumn{1}{c|}{18.52\%} &
\multicolumn{1}{c|}{14.20\%} &
\multicolumn{1}{c|}{13.37\%} &
\multicolumn{1}{c|}{18.09\%} &
\multicolumn{1}{c|}{13.21\%} &
\multicolumn{1}{c|}{17.28\%} &
\multicolumn{1}{c|}{14.02\%} &
\multicolumn{1}{c|}{15.45\%} &
\multicolumn{1}{c|}{15.45\%} &
\multicolumn{1}{c|}{22.56\%} &
\multicolumn{1}{c|}{25.00\%} &
\multicolumn{1}{c|}{24.90\%} &
\multicolumn{1}{c|}{23.04\%} &
\multicolumn{1}{c|}{19.55\%}
  \\ \cline{3-19} 
 
& 
  \multicolumn{1}{|c|}{\cellcolor{incorrectres}\textbf{Reasoning}} 
  &\cellcolor{incorrectout}\textbf{\begin{tabular}[c]{@{}c@{}}Incorrect Output\end{tabular}} 
  &\multicolumn{1}{c|}{3.09\%} 
  &\multicolumn{1}{c|}{0.00\%} 
  &\multicolumn{1}{c|}{1.23\%} 
  &\multicolumn{1}{c|}{1.23\%} 
  &\multicolumn{1}{c|}{0.00\%}
  &\multicolumn{1}{c|}{1.12\%} 
  &\multicolumn{1}{c|}{0.00\%}
  &\multicolumn{1}{c|}{0.41\%} 
  &\multicolumn{1}{c|}{0.21\%}
  &\multicolumn{1}{c|}{0.20\%}
  &\multicolumn{1}{c|}{0.61\%}
  &\multicolumn{1}{c|}{1.01\%}
  &\multicolumn{1}{c|}{0.00\%}
  &\multicolumn{1}{c|}{0.00\%}
  &\multicolumn{1}{c|}{0.00\%}
  &\multicolumn{1}{c|}{0.00\%}
  \\ \hline
  
\end{tabular}
}
\vspace{-10pt}
\end{table*}
\vspace{-5pt}
\section{Evaluation}
\label{sec:evaluation}

We investigate the following research questions in the paper\footnote{\textcolor{\rebuttalColor}{\approach artifacts~\cite{website} includes the results of additional experiments.}}:

    \textbf{RQ1. Performance in \approach.} To what extent can LLMs simulate the program execution?

    \textbf{RQ2. Reasoning Consistency Across Tests.} Can LLMs consistently simulate different execution paths of the same program?

    \textbf{RQ3. Diagnostic Analysis.} At which program points are LLMs more likely to start diverging from the real execution? What are the root causes for incorrect or suspiciously correct output predictions?

    \textbf{RQ4. \approach and Bug-related Tasks.} Can LLM reason about execution when solving programming tasks requiring control and data-flow awareness? Can \approach help vet the suspicious success of LLMs in those tasks?

    \textbf{RQ5. Comparison With Other Approaches.} How does \approach compare with existing code reasoning techniques?


\vspace{-8pt}
\subsection{Experimental Setup}
\label{subsec:setup}

\textbf{Subject LLMs.}
We select \textcolor{\rebuttalColor}{$16$} pre-trained or instruction-tuned models of different sizes, covering both general-purpose and Code LLMs: 
\textcolor{\rebuttalColor}{\omini~\cite{o4mini2025}, Gemini-2.5-Pro~\cite{gemini25pro2025}, DeepSeek-R1~\cite{guo2025deepseek},} 
\gptf~\cite{achiam2023gpt}, \gemini~\cite{team2023gemini}, CodeLlama~\cite{roziere2023code} (Base-7b, Instruct-7b, Base-13b, Instruct-13b, Instruc-34b), DeepSeekCoder~\cite{bi2024deepseek} (Instruct-6.7b, Base-6.7b, Instruct-33b), Magicoder-S-6.7b~\cite{wei2023magicoder}, SemCoder-S~\cite{ding2024semcoder}, and StarCoder2-15b~\cite{lozhkov2024starcoder}. We downloaded the open-access LLMs from HuggingFace~\cite{huggingface} and followed the best prompting practices from their official documents to ensure proper evaluation. Our experimental setting enforces a temperature of zero for all the non-reasoning models to ensure the reproducibility of results. For other parameters, we use the default setting of each model.

\noindent \textbf{Subject Programs.}
We evaluate LLMs on HumanEval~\cite{humaneval}, the most widely used dataset of $164$ Python programs. The rationale is two-fold: first, evaluating the most recent Code LLMs under different programming tasks demonstrates a great performance on HumanEval. As we will show, such an outstanding performance is not necessarily due to code understanding, and many of them involve incorrect and unreasonable CoT shortcuts (\S\ref{subsec:rq4}); hence, it should not be considered a victory. Furthermore, HumanEval comes with extra artifacts, i.e., human-written bugs, which are required for RQ4 to compare \approach and bug-related tasks. Due to resource constraints, we sampled three tests for each HumanEval program to use in the evaluation. When sampling, we controlled for the selection of at least two tests of different prime paths. 

\vspace{-10pt}
\subsection{RQ1. Performance in CES}
\label{subsec:rq1}

Table \ref{table:rq1} shows the success in coherent reasoning and final output prediction. Detailed results about the misprediction of other properties are in RQ2. To better analyze the results in this and subsequent research questions, we also break down the results into four groups of programs: programs with conditional statements only (CO), programs with loop statements only (LO), programs with both loops and conditional statements (LC), and simple programs with no looping or branching (Others). For the last category, none of the coherency rules (\S \ref{sec:valid-invalid}) applies, and hence, we only report output prediction results. When reporting the aggregated result (last four rows), we count those cases under \textit{Coherent Reasoning} rows. We categorize our observations based on the coherency of the reasoning.

\vspace{-5pt}
\subsubsection{Coherent Reasoning Process.} LLMs are more likely to simulate the execution with coherent (\textcolor{\rebuttalColor}{$81.42\%$}) than incoherent reasoning (\textcolor{\rebuttalColor}{$18.58\%$}), but they mostly yield incorrect output predictions. 
    
\hspace{-4pt} $\bullet$ \hspace{1pt} \textbf{Impact of size.} Within the family of models, LLMs with more parameters always outperform smaller ones on correct output prediction: the performance improves from $18.40\%$ (\codellama-Instruct-7b) to $25.17\%$ (\codellama-Instruct-34b) and from $31.51\%$ (\deepseek-Instruct-6.7b) to $39.43\%$ (\deepseek-Instruct-33b). Between the models of different sizes, bigger models outperform smaller ones, except \deepseek-Instruct-34b slightly outperforms \gemini in correct output prediction. 

\hspace{-4pt} $\bullet$ \hspace{1pt} \textbf{Impact of instruction-tuning.} Instruction-tuning slightly improves the LLMs' performance: for \codellama-7b, \codellama-13b, and \deepseek-6.7b, the instruction-tuned version outperforms the base with the margin of $1.96\%$, $1.65\%$,  and $3.47\%$, respectively. This is because the instruction-tuned versions follow prompt instructions better, which is important since \approach is a complex task. For \sem-S\cite{ding2024semcoder}, fine-tuned on \deepseek-Base-6.7b with \emph{execution data}, the improvement is $6.92\%$. \sem-S also outperforms instruction-tuned models of the same size or even bigger, demonstrating the impact of execution-aware fine-tuning in better code reasoning\footnote{The authors have decontaminated the dataset from HumanEval problems \textcolor{\MRColor}{ to ensure the improvement doesn't come from accessing the test set}.}.
\textcolor{\rebuttalColor}{Reasoning-based LLMs (\omini, Gemini-2.5-Pro, and \dsr) surpass \gptf and \gemini by $18.67\%$, on average. This suggests that LLMs incorporate internal reasoning, albeit different, into new reasoning tasks like \approach.} 
    
\hspace{-4pt} $\bullet$ \hspace{1pt} \textbf{Impact of code constructs.} LLMs struggle with loops (row \emph{LO}) and complex programs with different code constructs (row \emph{LC}), as the average correct output prediction for programs in these categories drops significantly compared to \emph{CO} programs. Given the simplicity of programs in HumanEval, LLMs are likely to struggle more in the execution simulation of real-world programs.

\subsubsection{Incoherent Reasoning Process} Surprisingly, LLMs with good performance in coherent reasoning and correct output prediction also result in more incoherent reasoning. Among all the models, \gptf and \textcolor{\rebuttalColor}{\dsr} generate more incoherent reasoning cases, $25\%$ and \textcolor{\rebuttalColor}{$24.90\%$}, respectively. In a similar trend, the base models that previously underperformed instruction-tuned models generate less suspiciously correct outputs. As we will discuss later with more details and in-depth analysis (\S\ref{subsec:rq3}), we speculate this is due to the interference of natural language reasoning and code reasoning. That is, instruction-tuned models better aligned with natural language instructions may \emph{override} code reasoning with natural language shortcuts or hallucinations. The same observation holds for \sem, which generates more suspiciously correct predictions compared to \deepseek-Base-6.7b. While \sem incorporates execution information such as coverage, orders, and program states, it uses natural language monologues for instruction tuning. As a result, it falls into the same trap and hallucinates, resulting in incoherent reasoning. We believe these results should initiate rethinking pre-training or fine-tuning strategies for more realistic Code LLMs that can better reason about code.

\vspace{-5pt}
\subsection{RQ2. Reasoning Consistency Across Tests}
\label{rq:reasoning-consistency}
\begin{figure}[t]
    \centering
    \includegraphics[width=0.85\linewidth]{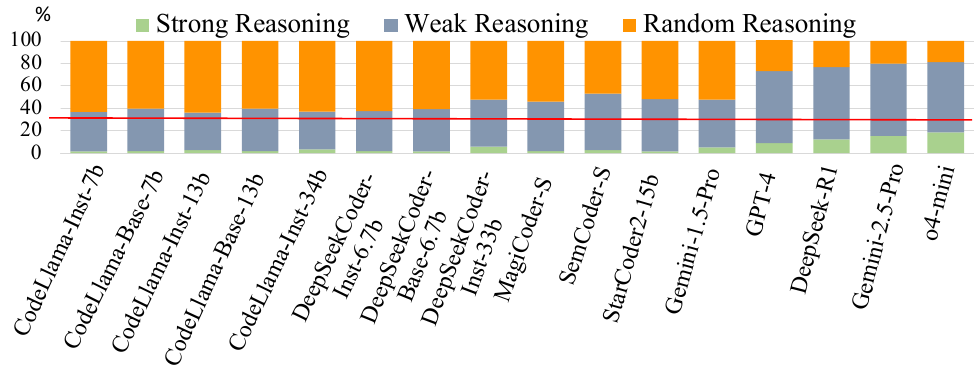}
    \vspace{-8pt}
    \caption{\textcolor{\rebuttalColor}{The reasoning consistency of LLMs across three sampled tests on \heval programs}}
    \label{fig:consistency}
    \vspace{-15pt}
\end{figure}
To evaluate the extent to which LLMs can reason about different execution paths of the same program, we computed the percentage of programs they can \emph{strongly} (Formula~\ref{eq:strongr}), \emph{weakly} (Formula~\ref{eq:weakr}), or \emph{randomly} simulate executions. Figure~\ref{fig:consistency} shows the result of this study. It is worth noting that not all the \heval programs have tests with different prime path coverage, due to incompleteness of test suite or being simple with no looping or branching ($53$ programs under \emph{Others} category of Table~\ref{table:rq1}). The red line in Figure~\ref{fig:consistency} marks the percentage of programs with at least two tests with different prime path coverage ($52$ out of $164$ programs ($31.71\%$)). 

These results show that LLMs, in \textcolor{\rebuttalColor}{$5.76\%$, $45.37\%$, and $48.87\%$} of programs, on average, \textcolor{\MRColor}{strongly, weakly, and randomly} simulate the execution across different tests. When considering the $52$ programs with at least two tests with different prime path coverage, LLMs strongly, weakly, and randomly reason about  \textcolor{\rebuttalColor}{$17.88\%$, $38.70\%$, and $43.42\%$} of them. The notion of self-consistency among different tests is simpler than self-consistency among different tasks~\cite{min2023beyond,huang2023enhancing} since the task here is not changing. Yet, most LLMs cannot consistently simulate the same execution paths under different inputs.

\begin{figure}[t]
 \centering
 \includegraphics[width=0.45\textwidth]{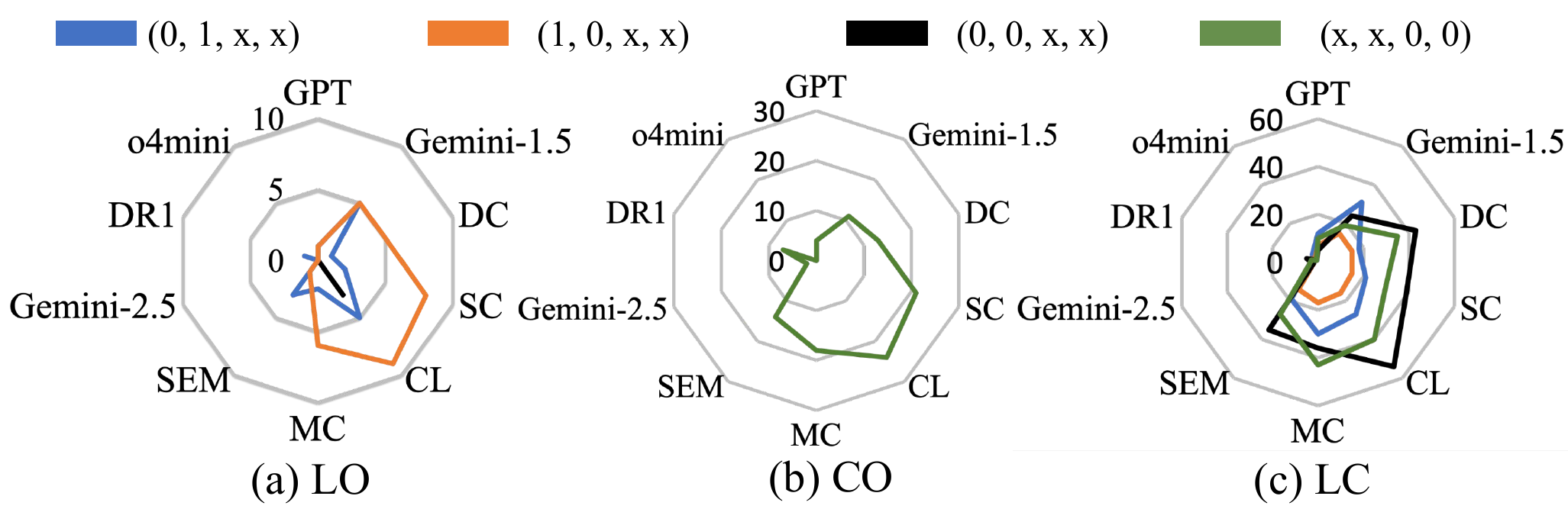}
 \vspace{-8pt}
 \caption{\textcolor{\rebuttalColor}{Comparison of simulation divergence locations for top ten models.
 $\langle V_l, I_l, P_c, B_c \rangle$ denotes location of divergence. The notion of $x$ implies the property does not apply to the location. GPT: (GPT-4), CL: (CodeLlama), DS (DeepSeekCoder), MC (MagiCoder), SEM (SemCoder), SC (StarCoder2), Gemini-1.5 (Gemini-1.5-Pro), Gemini-2.5 (Gemini-2.5-Pro), DR1 (DeepSeek-R1), and o4mini(o4-mini)}}
\vspace{-15pt}
\label{fig:radar-incorrect-output}
\end{figure}
\vspace{-8pt}
\subsection{RQ3. Diagnostic Analysis}
\label{subsec:rq3}

\subsubsection{Simulation Divergence.} For incorrect output predictions, \approach can automatically determine where the execution simulations started to diverge and propagate the incorrect state. Figure~\ref{fig:radar-incorrect-output} compares the simulation divergence locations across different categories of programs for the seven models (one model per each family) from Table~\ref{table:rq1}. We represent locations, which are either $S_{loop}$ or $C_{condition}$, with quadruples $\langle V_{l_j}, I_{l_j}, P_{c_j}, B_{c_j} \rangle$, corresponding to program properties (\S\ref{subsec:properties}). The $1$ or $0$ for an element indicates whether the property is correctly predicted. When a property does not apply to the location, the element value would be `$x$': for $S_{loop}$, only the first two elements are applicable; for $S_{condition}$, only the last two elements are applicable, and they can only be $0$, i.e., $(x,x,0,0)$. Correct values of them ($(x,x,1,1)$) that yield correct prediction violate coherency Rule 3 (Formula~\ref{eq:invalid3}), and a mismatch between their values violates coherency Rule 2 (Formula~\ref{eq:invalid2}). In LO programs, mispredicting loop iterables---$(1,0,x,x)$---is the most common reason to initiate the divergence. For CO programs, mispredicting predicates and, hence, branches---$(x,x,0,0)$---results in divergence. In LC programs, LLMs tend to mispredict both loop properties more---$(0,0,x,x)$, resulting in divergence from ground truth. The polygons inside the spider charts are mostly non-convex and overlapping, demonstrating the different behaviors of the LLMs. We manually investigated these cases to understand \emph{why} the predictions differed from ground-truth. Our investigations reveal several shortcomings of the subject LLMs:

\textcolor{\rebuttalColor}{LLMs fail to track the loop iterable whose value is dynamically changing inside the loop. In the example below (\heval/$13$), in the first loop iteration, the LLM incorrectly calculates the new value of the loop variable \texttt{\small{b}}, which propagates through all subsequent iterations and results in incorrect output prediction.}
\begin{lstlisting}[style=stylewithcommentPy]
def greatest_common_divisor(a: int, b: int) -> int:
    while b: ~##[STATE]b=[60,144,0][/STATE]~
        a, b = b, a % b
    return a

@@Input:@@ greatest_common_divisor(144, 60)
@@Ground Truth:@@ b=[60,24,12,0],output=12
@@Predicted Output (DeepSeekCoder-Inst-33b):@@ 60
\end{lstlisting}

\textcolor{\rebuttalColor}{LLMs may struggle with compound properties, e.g., conditional statements with multiple sub-predicates. Below (\heval/$57$), the \texttt{\small{if}} condition predicate constitutes two sub-predicates \texttt{\small{l==sorted(l)}} and \texttt{\small{l==sorted(l,reverse=True)}}. The LLM correctly predicts the first sub-predicates, but mispredicts the second, resulting in the misprediction of combined predicate and final output.}

\begin{lstlisting}[style=stylewithcommentPy]
def monotonic(l: list):
    if l==sorted(l) or l==sorted(l,reverse=True):
    ##[CONDITION](l==sorted(l))=False[/CONDITION]
    ~##[CONDITION](l==sorted(l,reverse=True))=False[/CONDITION]~
    ~##[CONDITION](l==sorted(l) or l==sorted(l,reverse=True))=False~
    ~[/CONDITION] 
    ##[BRANCH]taken=[N][/BRANCH]~
        return True
    return False
    
@@Input:@@ monotonic([4,1,0,-10])
@@Ground Truth:@@ (l==|sorted|(l))=False,(l==|sorted|(l,reverse=True))=True, 
(l==|sorted|(l) |or| l==|sorted|(l,reverse=True))=True,|if| branch=[Y],output=True
@@Predicted Output (Gemini-1.5-Pro):@@ False
\end{lstlisting}

\textcolor{\rebuttalColor}{Nested constructs can challenge LLMs more with execution simulation. The example below (\heval/$73$) shows a program with a nested construct: an \texttt{\small{if}} statement inside a \texttt{\small{for}} loop. The LLM mispredicts both the loop variable, \texttt{\small{range(len(arr)//2)}}, and the loop iterable, \texttt{\small{i}}. As a result, the execution simulation iterates over the loop body one additional time, causing mispredictions in the conditional predicates and the loop’s internal branch.}

\begin{lstlisting}[style=stylewithcommentPy]
def smallest_change(arr):
    ans = 0
    for i in range(len(arr) // 2): ~##[STATE]i=[0,1,2,3][/STATE]~ 
    ~##[STATE]range((len(arr)//2))=[0,1,2,3][/STATE]~ 
    ~##[STATE](len(arr)//2)=[4][/STATE]~
        if arr[i] != arr[len(arr) - i - 1]: 
        ~##[CONDITION](arr[i]!=arr[len(arr)-i-1])=[True,False,False
        ,False][/CONDITION] ##[BRANCH]taken=[Y,N,N,N][/BRANCH]~
            ans += 1
    return ans
    
@@Input:@@ smallest_change([1,2,3,4,3,2,2])
@@Ground Truth:@@ i=[0,1,2],|range((len(arr)//2))=[0,1,2]|,(|len|(arr)//2)=3,
arr[i]!=arr[|len|(arr)-i-1]=[True,False,False],|if| branch=[Y,N,N],
output=1
@@Predicted Output (CodeLlama-Inst-13b):@@ 2
\end{lstlisting}

\textcolor{\rebuttalColor}{LLMs may also struggle to reason about complex arithmetic/logic operations or mispredict the return value of API calls. Such cases, if they happen at the looping, branching, or return statements, \approach can automatically pinpoint them and accurately determine the point of simulation divergence. Otherwise, \approach determines the first location it observes the propagation of the misprediction.}

\subsubsection{Suspiciously Correct Output Prediction.} \approach can automatically detect suspiciously correct output predictions: To that end, it first recognizes incoherent execution simulation cases using the coherency rules (Formulas~\ref{eq:invalid1}--\ref{eq:invalid3}), and if the output prediction is correct, \approach marks them as suspicious. \approach, overall, detected $361$ cases of suspiciously correct output predictions (\emph{Incoherent Reasoning/Correct output} rows in Table~\ref{table:rq1}). We sampled $64$ of such cases across all models and program categories to better understand the root causes of suspiciously correct output predictions. 

The most common culprit among the studied samples is the \emph{natural language shortcut}. In such cases, a monologue-style step-by-step thinking of code execution seemingly overrides loop or condition properties mispredictions and results in correct output prediction. \textcolor{\rebuttalColor}{An example of such cases is shown below, belonging to the execution simulation of \heval/$11$. The LLM mispredicts the conditional predicate inside the loop (\texttt{\small i==j}) in iterations $3$, $4$, and $6$. In the CoT reasoning, however, it looks at the method name (\texttt{\small string\_xor}), assumes that the method implements bitwise \texttt{\small XOR}ing, and uses this assumption to provide a correct output prediction.} 


\begin{lstlisting}[style=stylewithcommentPy]
def string_xor(a: str, b: str) -> str:
    def xor(i, j):
        if i == j: 
        ~##[CONDITION](i==j)=[True,False,False,False,False,False]
        [/CONDITION] ##[BRANCH]taken=[Y,N,N,N,N,N][/BRANCH]~
            return '0'
        else: ~##[BRANCH]taken=[N,Y,Y,Y,Y,Y][/BRANCH]~
            return '1'
    return ''.join(xor(x, y) for x, y in zip(a, b))

@@Input:@@ string_xor('111000','101010')
@@Ground Truth:@@ (i==j)=[True,False,True,True,False,True],|if| branch=
[Y,N,N,N,N.N],|else| branch=[N,Y,Y,Y,Y],output=|'010010'|
@@Predicted output (Gemini-1.5-Pro):@@ |'010010'|
@@CoT:@@The function `string_xor` takes two binary strings as |input| |and| 
returns a new binary string that |is| the result of XORing them. 
@@If `a` is '111000' and `b` is '101010', then the output of the
@@ @@`string_xor` function will be '010010'.@@
\end{lstlisting}

We also observed cases where simulation execution and CoT reasoning were incorrect, yet the LLM hallucinated the correct output. \textcolor{\rebuttalColor}{Figure~\ref{fig:illistrative-example-2} shows an example of such cases, where the LLM, magically, responds with a correct output, despite incorrect simulation execution and CoT reasoning. Most of the other similar cases happened in programs with boolean return statements; thereby, we speculate that correct output predictions are, in fact, lucky hallucinations.}

\begin{table*}[t]
\caption{The performance of LLMs in \approach and bug prediction, localization, repair tasks.}
\vspace{-8pt}
\label{table:repair}
\scalebox{0.65}{
\begin{tabular}{|l|c|c|c|c|c|c|c|c|c|c|}
\hline
\textbf{} &
  \textbf{\begin{tabular}[c]{@{}c@{}}\codellama-Inst-34b\end{tabular}} &
  \textbf{\begin{tabular}[c]{@{}c@{}}\deepseek-Inst-33b\end{tabular}} &
  \textbf{\magic-S} &
  \textbf{\sem-S} &
  \textbf{\begin{tabular}[c]{@{}c@{}}\starT\end{tabular}} &
  \textbf{\gemini} &
  \textbf{\gptf} &
  \textbf{\textcolor{\rebuttalColor}{DeepSeek-R1}} &
  \textbf{\textcolor{\rebuttalColor}{Gemini-2.5-Pro}}&
  \textbf{\textcolor{\rebuttalColor}{o4-mini}}
  \\ \hline
\textcolor{black}{\textbf{Bug Prediction}}   & \textcolor{black}{18.75\%}  & \textcolor{black}{28.75}\% & \textcolor{black}{26.25\%} & \textcolor{black}{18.13\%} & \textcolor{black}{15.00\%} & \textcolor{black}{92.50\%} & \textcolor{black}{88.75\%} &
\textcolor{black}{83.13\%} &
\textcolor{black}{84.28\%}&
\textcolor{black}{95.00\%}
\\ \hline

\textcolor{black}{\textbf{Bug Localization}} & \textcolor{black}{41.88\%} & \textcolor{black}{40.63\%} & \textcolor{black}{24.38\%} & \textcolor{black}{44.38\%} & \textcolor{black}{33.13\%} & \textcolor{black}{72.50\%} & \textcolor{black}{71.25\%} &
\textcolor{black}{81.25\%} &
\textcolor{black}{81.25\%}&
\textcolor{black}{80.00\%}
\\ \hline
\textcolor{black}{\textbf{Bug Repair}}       & \textcolor{black}{42.50\%} & \textcolor{black}{76.25\%} & \textcolor{black}{69.38\%} & \textcolor{black}{74.38\%} & \textcolor{black}{60.00\%} & \textcolor{black}{90.00\%} & \textcolor{black}{93.13\%}&
\textcolor{black}{95.63\%} &
\textcolor{black}{91.25\%}&
\textcolor{black}{96.88\%}
\\ \hline

\textcolor{black}{\textbf{CES}}       & \textcolor{black}{18.13\%} & \textcolor{black}{27.50\%} & \textcolor{black}{20.61\%} & \textcolor{black}{23.75\%} & \textcolor{black}{25.00\%} & \textcolor{black}{32.50\%} & \textcolor{black}{55.00\%} &
\textcolor{black}{70.00\%} &
\textcolor{black}{77.50\%} &
\textcolor{black}{75.00\%}
\\ \hline

\end{tabular}
}
\vspace{-8pt}
\end{table*}

\vspace{-8pt}
\subsection{RQ4. \approach and Bug-Related Tasks}
\label{subsec:rq4}

\begin{figure}[t]
    \centering
    \includegraphics[width=0.99\linewidth]{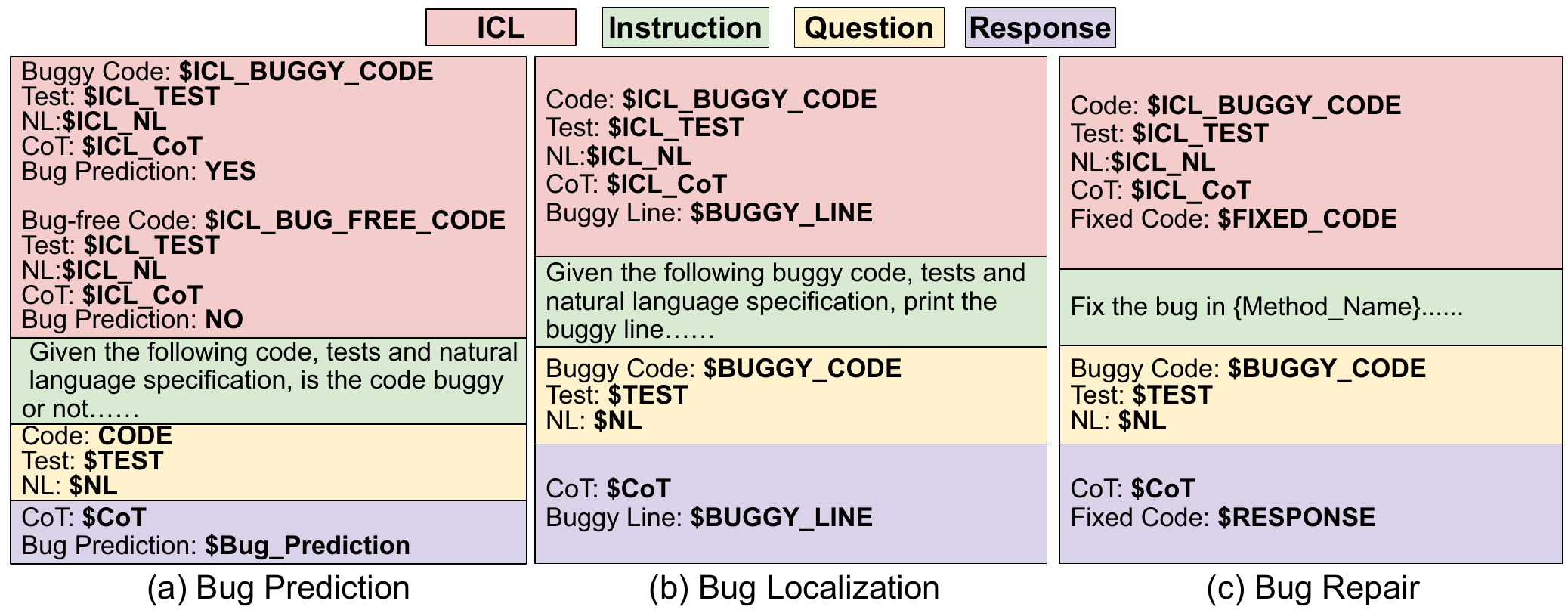}
    \vspace{-8pt}
    \caption{Prompt template of bug-related tasks. ICL stands for in-context learning}
    \vspace{-15pt}
    \label{fig:prompt_bug_tasks}
\end{figure}

LLMs should incorporate their knowledge of programming languages and examples they have seen to solve programming tasks. Otherwise, one cannot expect them to generalize to different tasks or perform reliably in real-world settings. Concerning programming tasks, specifically bug-related tasks, we expect them to predict, localize, or repair bugs based on program analysis. Assuming execution simulation as a required step for program analysis, we can define the following expectations: (1) if an LLM can correctly simulate an execution path, it is more likely to detect, localize, and repair a bug in that specific execution path. Similarly, (2) if a model cannot correctly simulate an execution path, it either fails in bug-related tasks or the success is based on other factors, e.g., pattern matching, reasoning shortcuts, hallucination, or data leakage.
\begin{figure*}[t]
    \centering
    \includegraphics[width=0.88\linewidth]{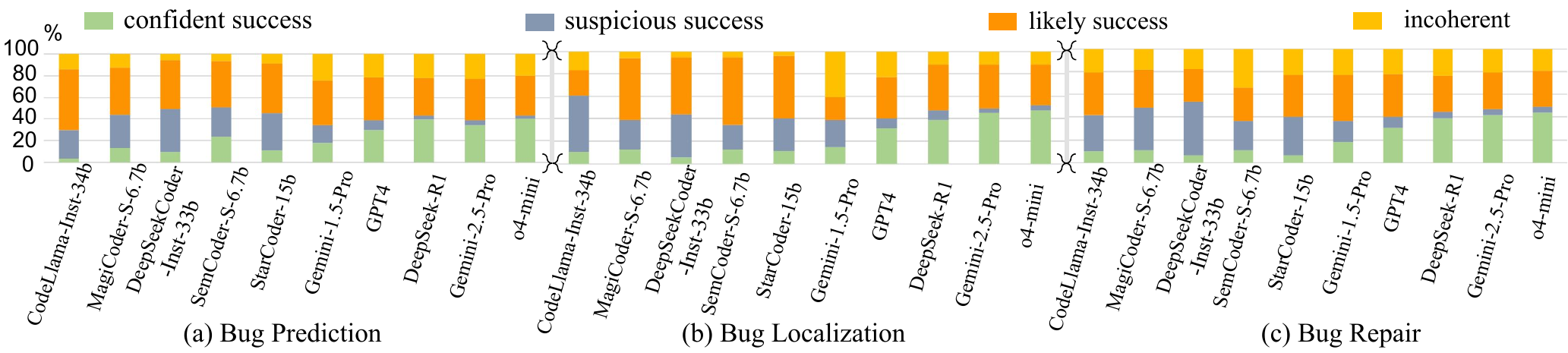}
    \vspace{-8pt}
    \caption{\textcolor{\rebuttalColor}{Using \approach to vet the success of LLMs in bug prediction (a), localization (b), and repair (c) tasks}}
    \label{fig:explainability-bug-tasks}
    \vspace{-13pt}
\end{figure*}
To investigate whether LLMs meet these two expectations, we evaluated the subject LLMs with three prominent bug-related tasks: bug prediction, localization, and repair. For this experiment, we used HumanEvalPack~\cite{muennighoff2023octopack}, a buggy version of \heval with the bugs developed and injected by humans. The dataset comes with a set of regression and error-revealing tests. 

We picked a model from each family and prompted them with the templates shown in Figure \ref{fig:prompt_bug_tasks}-a--\ref{fig:prompt_bug_tasks}-c. The prompt instructs LLMs for individual tasks with in-context examples. For the bug prediction, the prompt provides one buggy and one correct example. For bug localization and prediction, it provides only the buggy example. Prompt templates also include the docstring and an error-revealing test. We instruct LLMs to explain the execution process of the program in the CoT. The rationale for this design decision is three-fold: (1) for the sake of fairness, we want the prompts between \approach and these tasks to be similar; (2) we wanted to see if instructed, LLMs can incorporate simulation execution into decision-making; and (3) we can check the CoT for further analysis of the results. We also prompted selected LLMs for execution simulation of the error-revealing tests used in bug-related tasks.

Table~\ref{table:repair} shows the performance of models in these four tasks, and we also analyze the overlap between the successful cases of each task per individual LLM (Figure~\ref{fig:rq4-venn}). For stronger models, although small, we observe overlaps between the success cases. \textcolor{\rebuttalColor}{For example, \omini and Gemini-2.5-Pro can succeed in all four tasks on the $58.13\%$ and $48.13\%$} of \heval programs. For some weaker models, the overlap is negligible, close to zero. The overlap, even among the bug-related tasks, is minimal, which motivates deeper investigation into how LLMs succeed in these tasks. 

The design of \approach provides us with a robust systematic analysis to vet the suspicious success of LLMs in bug-related tasks. If the execution simulation is \emph{coherent} for those cases, depending on the result of output prediction, we can perform the following analyses:

\begin{figure}[t]
    \centering
    \vspace{4pt}
    \includegraphics[width=\linewidth]{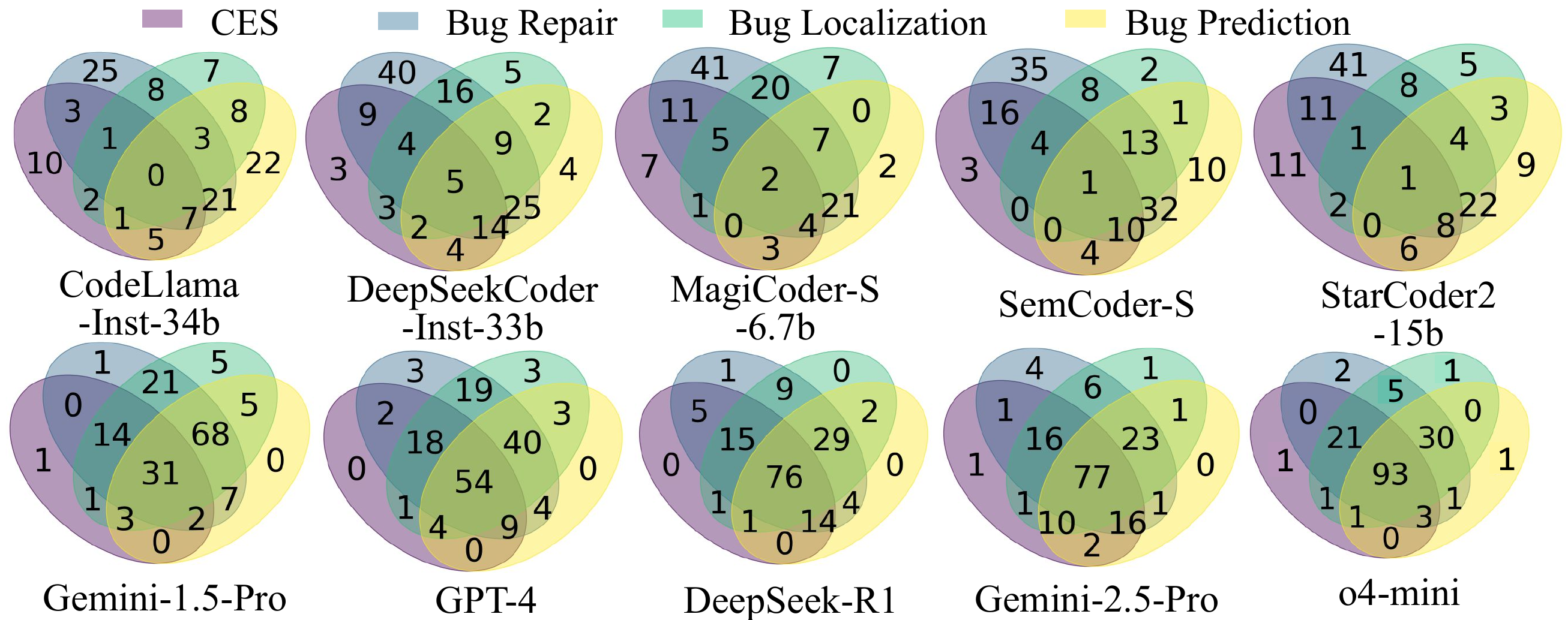}
    \vspace{-8pt}
    \caption{\textcolor{\rebuttalColor}{Overlap between success cases in bug-related tasks and \approach for ten representative subject LLMs}}
    \label{fig:rq4-venn}
    \vspace{-10pt}
\end{figure}

\hspace{-4pt} $\bullet$ \hspace{1pt} \emph{Correct output prediction.} For the programs that the model succeeds in \approach and the given bug-related task, the LLM likely has incorporated execution reasoning into account for predicting, localizing, or repairing the bug. We refer to these cases as \emph{confident success} for bug-related tasks.

\begin{figure*}[t]
    \centering
    \includegraphics[width=0.84\linewidth]{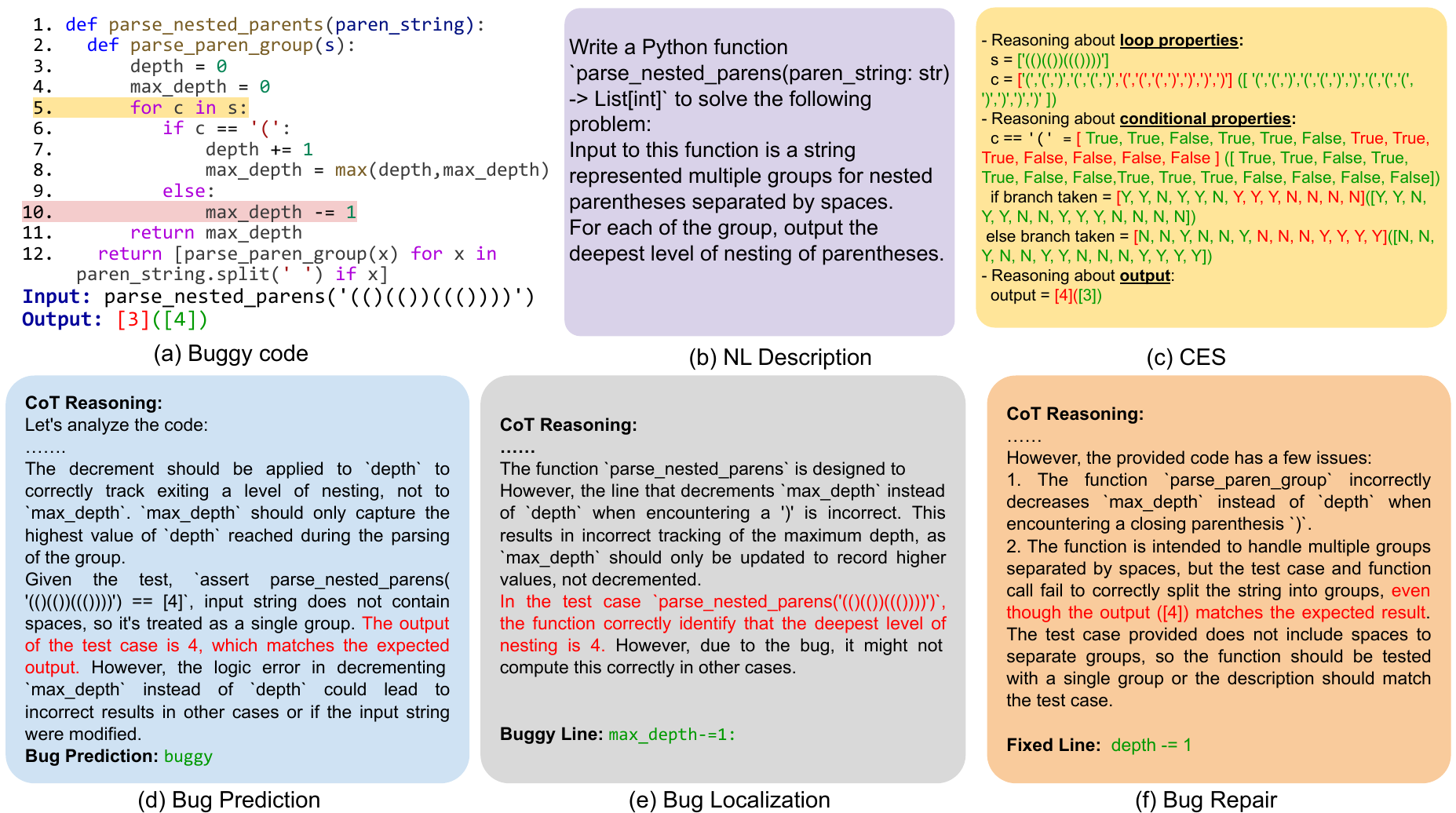}
    \vspace{-8pt}
    \caption{An example showcasing incorrect output prediction in \approach (c), and correct bug prediction (d), localization (e), and repair (f) by \gptf on HumanEval/6 (a). The specification for the functionality of HumanEval/6 is shown in (b)}
    \label{fig:sus}
    \vspace{-10pt}
\end{figure*}

\begin{figure*}[t]
    \centering
    \includegraphics[width=0.84\linewidth]{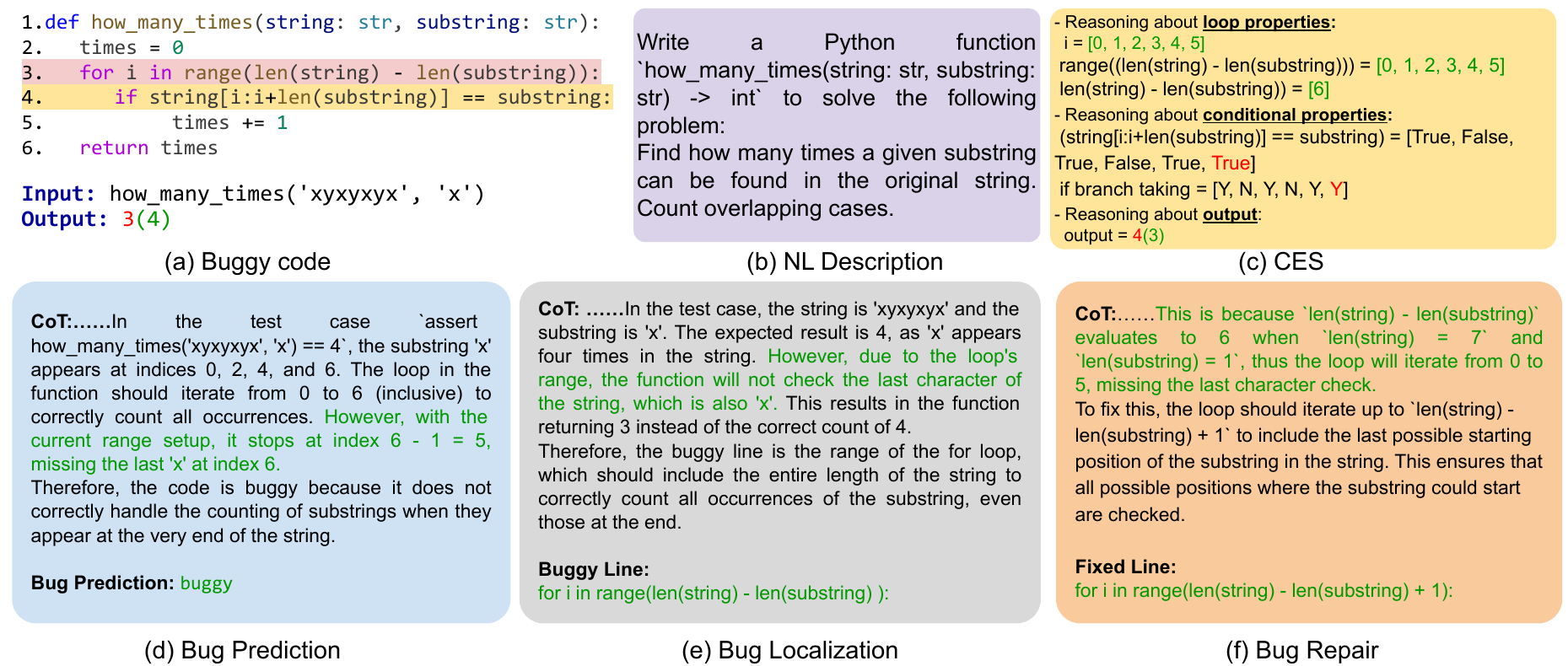}
    \vspace{-8pt}
    \caption{An example showcasing incorrect output prediction in \approach (c), and correct bug prediction (d), localization (e), and repair (f) by \gemini on HumanEval/18}
    \label{fig:less-sus}
    \vspace{-10pt}
\end{figure*}

\hspace{-4pt} $\bullet$ \hspace{1pt} \emph{Incorrect output prediction.} For the programs that LLM fails to correctly simulate the execution of error-revealing tests, its success in bug-related tasks could be due to pattern matching, lucky hallucination, natural-language shortcuts, or data leakage. The only exception here is that if \emph{divergence location} is \emph{after} the bug location, the LLM could have incorporated execution simulation and observed the buggy behavior. However, it could not correctly continue the simulation until the end. We refer to these cases as \emph{likely success} for bug-related tasks. For the cases that divergence location is \textcolor{\MRColor}{\emph{before} the bug location}, we refer to them as \emph{suspicious success}. 

Figures~\ref{fig:explainability-bug-tasks}-a--\ref{fig:explainability-bug-tasks}-c show the result of automated analysis and vetting on the successful bug-related tasks from the previous experiment. Each bar in this figure represents a given model's success cases per the specific tasks, divided into four categories according to \approach results for the same programs. On average, \textcolor{\rebuttalColor}{$22.54\%$, $39.25\%$, and $20.75\%$} of the correct predictions on programming tasks are \correct, \lesssus, and \sus, respectively. Manual investigation of the \lesssus cases show that LLMs, in their CoT, indeed tried to incorporate simulation execution into the problem-solving process. On the other hand,  there are lucky hallucinations and shortcuts galore in the \sus cases. These results confirm that \approach, besides evaluating code reasoning abilities of LLMs, can serve as an effective and systematic vetting technique to rule out suspicious correctness in bug-related tasks.  

Figure~\ref{fig:sus} shows a \emph{\sus} case, demonstrating the performance of \gptf on HumanEval/6 under four studied tasks. The buggy code is shown in Figure~\ref{fig:sus}-a, which for the test input \texttt{\small{`(()(())((())))'}}, is expected to return \texttt{\small{$[4]$}}. Due to the bug in line $10$, however, the code returns \texttt{$[3]$}. From Figures~\ref{fig:sus}-d to \ref{fig:sus}-f, we can see that \gptf is capable of predicting, localizing, and fixing the bug. Figure \ref{fig:sus}-(c) also shows that \gptf diverges from the ground truth execution since line $5$ (before the bug in line $10$), making it unlikely that the model incorporated the simulation execution into problem-solving for bug-related tasks. Looking at the CoT of bug-related tasks, we observe no notable reasoning concerning execution simulation. In fact, the CoT only explains the high level of the code, including incorrect assumptions that the test output matches the expected one on the buggy code. This example indicates that LLMs neglected the test information and solely relied on the natural language specification to succeed in bug-related tasks. This, by itself and specific to this example, may not be problematic. However, it showcases a concerning threat to generalize abilities to case specifications that do not exist, and LLMs may only be provided with tests to predict, localize, or repair the bug. Such cases happen in practice often, if not always. 

Figure ~\ref{fig:less-sus} shows the performance of \gemini on \heval/18 under the four studied tasks. The bug location is in the loop iterable of line $3$ (Figure~\ref{fig:less-sus}-a). In the correct code, the loop iterable should be \texttt{\small{len(string)-len(substring)+1}}, while in the buggy code, it is \texttt{\small{len(string)-len(substring)}}. As a result, given the inputs \texttt{\small{'\`xyxyxyx',`x'}}, the buggy code returns $3$ instead of $4$. The divergence location is at line $4$, after the buggy line, due to the misprediction of the conditional predicate and branch. Looking at the \approach reasoning (Figure~\ref{fig:less-sus}-c), \gemini correctly predicts the number of iterations but mispredicts the conditional predicate at the last iteration. Looking at the CoT of bug-related tasks, we can see that \gemini simulates the execution of the buggy code and correctly identifies where the execution does not match the specification. In software testing, when a test finds a bug, its execution can terminate. The same concept applies here: as long as execution simulation propagates the flow of inputs to buggy location, the rest of the execution simulation does not matter much.

\vspace{-8pt}
\subsection{RQ5. Comparison With Other Approaches}
\label{subsec:rq5}

We compare \approach with \reval and \codemind, both of which have evaluated code reasoning of several LLMs on the \heval programs. To ensure a fair comparison, we select $82$ programs (containing $240$ statements) used by all three approaches. \codemind and \reval provide the results for all $164$ and $82$ \heval programs, respectively. Results are reported for seven models in the overlap of \approach and at least one approach's experiments (Table \ref{table:comparison}). 

Output prediction under \approach is $42.98\%$ and $56.38\%$, lower than \codemind and \reval: \approach discards suspiciously correct predictions, which is not supported by other approaches. Without excluding the suspiciously correct predictions, the performance of models under \approach becomes closer to two other techniques (column \emph{+SC} in Table~\ref{table:comparison}). Also, \approach is more complex than the other two techniques. Although output prediction implicitly requires execution simulation, explicitly asking models to reason about intermediate program states complicates the task. This aligns with prior research~\cite{dziri2023faith}, which shows that LLMs struggle to handle compositional reasoning, i.e., decomposing problems into intermediate steps and synthesizing these steps into a precise answer.

\begin{table}[t]
\small
\caption{Comparison with \reval and \codemind. `NA' means the model's results are unavailable in the official artifact. \textbf{+SC} means suspiciously correct instances included.}
\vspace{-8pt}
\label{table:comparison}
\centering
\setlength{\tabcolsep}{2pt}
\resizebox{0.45\textwidth}{!}{
\begin{tabular}{|l|c|c|c|c|c|}
\hline
\multicolumn{1}{|c|}{\textbf{Model}} &  \multicolumn{4}{c|}{\textbf{Output Prediction}} & \textbf{Consistency} \\ 

 & \approach & +SC & \codemind  & \reval & \reval \\ \hline

\textbf{\codellama-Inst-7B}      
& 15.85\%  
& 42.68\%
& NA       
& 80.00\% 
& 9.50\%  \\ \hline

\textbf{\codellama-Base-7B} 
& 12.20\%
& 40.24\%
& NA       
& 77.56\% 
& 2.49\%  \\ \hline

\textbf{\codellama-Inst-13B}    
& 18.29\%   
& 45.12\%
& 75.85\% 
& 79.51\% 
& 3.36\%  \\ \hline

\textbf{\deepseek-Inst-6.7B} 
& 30.49\% 
& 57.31\%
& 63.41\% 
& NA 
& NA       \\ \hline

\textbf{\magic-S-6.7B}         
& 23.17\%   
& 47.56\%
& 74.39\% 
& 79.24\%
& 6.43\%  \\ \hline

\textbf{\starT-15B}            
& 34.15\%
& 64.63\%
& 71.95\% 
& 85.37\%
& 17.28\% \\ \hline

\textbf{\gptf}                
&  47.56\%
& 81.71\%
& 82.93\% 
& 87.80\% 
& 42.72\% \\ \hline
\end{tabular}
}
\vspace{-15pt}
\end{table}

\begin{figure}
     \centering
     \includegraphics[width=0.7\linewidth]{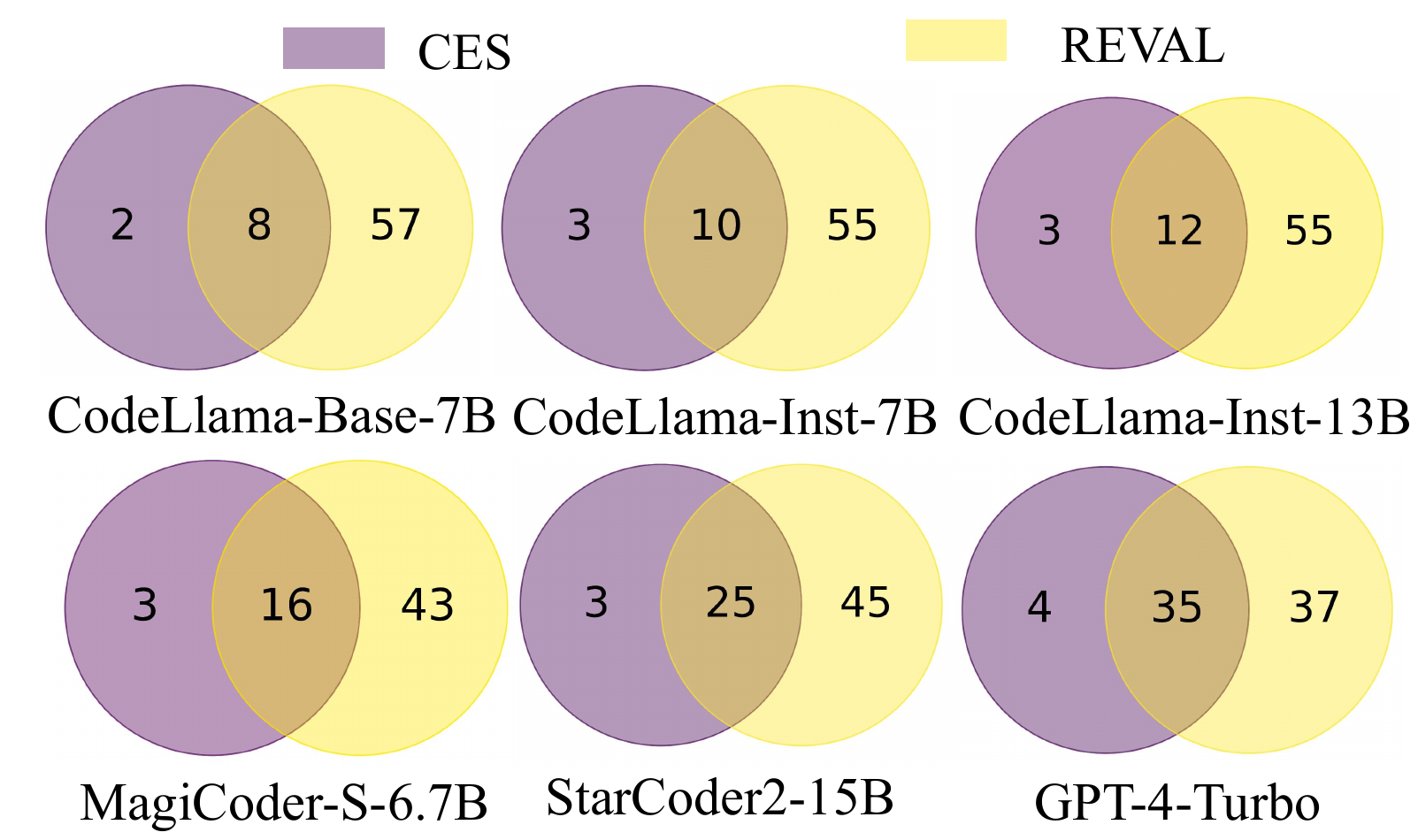}
     \vspace{-8pt}
     \caption{\textcolor{\MRColor}{Comparison between \approach and \reval}}
     \label{fig:reval_ces_comparison}
     \vspace{-15pt}
\end{figure}

A fairer comparison is between \approach output prediction and \reval consistency metric. \reval introduced the notion of \emph{incremental consistency} to account for separately prompting LLMs to perform a sequentially related task, i.e., program execution. In contrast, the \approach enables incremental consistency by design; it asks LLM to simulate program flow within one prompt. The average output prediction in \approach across all LLMs ($25.96$) is $1.9$ times higher than that of incremental consistency in \reval ($13.63$).
\textcolor{\MRColor}{Figure~\ref{fig:reval_ces_comparison} visualizes the overlap and differences between \approach and \reval in terms of successful output predictions. We observe that as a model's success rate in \approach increases, the intermediate reasoning helps its final output prediction more. For advanced LLMs, better evaluation tasks, such as \approach, are needed to evaluate their compositional reasoning.}

\vspace{-8pt}
\section{Related Works}
\label{sec:related_work}

A large body of work has evaluated LLMs on reasoning tasks across different modalities~\cite{deshpande2021rect,wu2023reasoning,miceli2023larger,bubeck2023sparks,wang2023mathcoder,imani2023mathprompter,luo2023wizardmath,huang2023lvlms,valmeekam2022large,min2023beyond}, including code reasoning \cite{roy2025codesense,wei2025equibench,xie2025core,liu2025real}. \crux and \codemind introduced early benchmarks for this area: \crux included input and output prediction tasks, while \codemind proposed three tasks—independent execution, dependent execution, and specification reasoning—to assess inductive code reasoning.

\reval took one step forward by including runtime behavior in the reasoning process, using four runtime prediction tasks: code coverage prediction, program state prediction, execution path prediction, and output prediction. Given that these tasks are inherently connected and sequential but are prompted separately, they proposed to measure the consistency of the models in performing these tasks sequentially as an ultimate evaluation metric for code reasoning. \coco~\cite{beger2025coconut}, challenges LLMs to generate a trace of line numbers executed by the program for a given set of inputs, which requires LLMs to reason about the program's control flow. 

\approach 
is the first technique to discuss a systematic and formal approach to identifying reasoning issues, e.g., hallucinations and reasoning shortcuts. The novel notion of consistency across tests offers an intuitive metric for challenging LLMs that incorporate execution reasoning into decision-making. \approach is the first to study the relationship between code reasoning and other tasks, providing valuable insights into how LLMs succeed in these tasks.

\vspace{-5pt}\section{Threats to Validity}
\label{sec:threats}

\noindent \textbf{External Validity.}
We selected LLMs of different sizes and training strategies, and our tool is publicly available to evaluate LLM code-execution simulations on other benchmarks. We expect code reasoning abilities to degrade as subjects become more complex~\cite{liu2025real}.

\vspace{2pt}
\noindent \textbf{Internal Validity.}
To account for nondeterminism, we repeated the experiments per program across multiple tests and evaluated the reasoning generalizability through a novel consistency metric. Results may be affected by implementation bugs. To address this threat, we thoroughly tested the pipeline (prompting models and processing responses) and cross-checked the results for correctness.

\vspace{2pt}
\noindent \textbf{Construct Validity.} 
\reval only studied a subset of \heval ($85$ out of $164$ programs), while we evaluated LLMs in all the programs. For the common programs, the selection of tests was also not consistent. To mitigate this threat, in comparing the \reval and \approach, we recompute the incremental consistency of \reval for the programs and test inputs that overlap with  \approach.

\vspace{-5pt}
\section{Concluding Remarks}
This paper proposes \approach, a code reasoning task that unifies the prediction of output and essential decision points. This unique design
enables \approach to be flow-sensitive and diagnostic. The novel reasoning consistency metric across multiple tests can assess the reasoning abilities of models in a spectrum: strong, weak, or random. Our results show that LLMs can simulate program execution. However, their reasoning is mostly random or, at best, weak. Our artifacts are publicly available for reproducibility and extension~\cite{website}. We plan to use the \approach for a large-scale evaluation of reasoning in real-world benchmarks such as SWE-Bench. 
Enabling the \approach for real-world projects requires non-trivial efforts to account for complex objects, making it beyond the scope of this paper. 

\vspace{-5pt}
\begin{acks}
This work is supported by NSF CCF 22-38045 CAR grant, IBM-Illinois Discovery Accelerator Institute, and Amazon-Illinois Center on AI for Interactive Conversational Experiences. 
\end{acks}



\bibliographystyle{ACM-Reference-Format}
\bibliography{sample-base}


\end{document}